\documentclass{ws-mplb}

\begin{document}

\markboth{Kakehashi et al.}{Momentum-Dependent Local Ansatz Approach to
Correlated Electron}

%
\catchline{}{}{}{}{}
%

\title{MOMENTUM-DEPENDENT LOCAL ANSATZ APPROACH TO CORRELATED 
ELECTRONS \footnote{Dedicated to the Late Professor Martin C. 
Gutzwiller. \ (To be published in Modern Physics Letters B.)}
}

\author{\footnotesize YOSHIRO KAKEHASHI${}^{\dagger}$ and SUMAL CHANDRA}

\address{Department of Physics, Faculty of Science, University of the
Ryukyus, Nishihara, Okinawa 903-0213, Japan \\
${}^{\dagger}$yok@sci.u-ryukyu.ac.jp}

\author{DERWYN ROWLANDS}

\address{Shanghai Key Laboratory of Special Artificial Microstructure
Materials and Technology, \\
School of Physics Science and Engineering, Tongji University, Shanghai
200092, China}

\author{M. ATIQUR R. PATOARY}

\address{Department of Physics, University of Rajshahi, Rajshahi 6205, 
Bangladesh}

\maketitle

\begin{history}
\received{27 June 2014}
\end{history}

\begin{abstract}
 The wavefunction method provides us with a useful tool to describe
 electron correlations in solids at the ground state.  In this paper 
 we review the recent development of the momentum-dependent
 local ansatz wavefunction (MLA).  It is constructed by taking into
 account two-particle excited states projected onto the local orbitals,
 and the momentum-dependent amplitudes of these states are chosen as
 variational parameters.  The MLA describes accurately correlated
 electron states from the weak to the intermediate Coulomb interaction
 regime in infinite dimensions, and works well even in the strongly
 correlated region by introducing a new starting wavefunction called the
 hybrid (HB) wavefunction.  The MLA-HB is therefore shown to overcome
 the limitation of the original local ansatz (LA) wavefunction as 
 well as the Gutzwiller wavefunction.  
 In particular, the calculated quasiparticle weight vs
 Coulomb interaction curve is shown to be close to that obtained by the
 numerical renormalization group approach.  It is also shown that 
the MLA is applicable to the first-principles Hamiltonian.
\end{abstract}


\section{Introduction}

Properties of solids are well-known to be determined by the quantum
mechanical motion of electrons.  The electrons move there in the
periodic potential and hence form energy bands, but they are also
influenced by the electron-electron interactions.  The latter effects
are usually taken into account as an effective mean-field potential.
The Hartree-Fock approximation is an approximate method to describe
the interactions with an effective potential.
The band theory based on the effective potential explains many aspects
of solids such as the cohesive properties, the Fermi surface in metals,
and optical properties of many metallic systems, especially when the
electron-electron interaction energy is small as compared with the
kinetic energy of band electrons~\cite{slater72,martin04}.

When electron-electron interactions become comparable to the kinetic
energy of electrons, various effects which cannot be explained by the
simple band picture appear.  These effects are called 
electron-correlation effects.  
Correlation effects originate in the quantum charge and spin 
fluctuations neglected in the mean field 
approximation~\cite{fulde95,fulde12}.

Correlated motion of electrons, for example, suppresses the electron
hopping in a solid, and yields the reduction of cohesive energy 
in 3d transition metals~\cite{friedel77}, 
the band narrowing, and the formation of 
a satellite peak in the X-ray photoemission spectroscopy (XPS) data 
of Ni~\cite{penn79,liebsch79}.
Strong mass enhancement of quasiparticle states is also a consequence of
the localization of electrons.  Further electron localization is known
to cause the Mott-type metal-insulator 
transition~\cite{mott90,gebhard97}.

In the magnetic materials, the quantum spin fluctuations are well known
to make the ferromagnetism at the ground state unstable, create the 
magnetic entropy at finite temperatures, and consequently much reduce 
the Curie temperatures~\cite{moriya85,kake04adv,kake12}.  
The long-range quantum spin fluctuations are
known to cause the spin wave excitations.
The high-temperature superconductivity in cuprates is also believed to
be caused by the antiferromagnetic spin fluctuations in the vicinity of
the Mott transition, which is missing in the band 
theory~\cite{anderson87,schrieffer07}.

The density functional theory for band structure calculations has
overcome some of these difficulties at the ground 
state using the effective potentials based on the local density
approximation (LDA) or the generalized gradient approximation 
(GGA)~\cite{kohn64,kohn65,martin04}. Most of the
problems mentioned above, however, cannot be understood without 
taking into account the correlated motion of electrons, and 
we need to develop the many-body theory of electron correlations.

Theoretical approaches to interacting electrons have been developed
since quantum mechanics was established.  The diagram technique is
the standard method to calculate the Green function and the free energy
for correlated electrons starting from non-interacting 
electrons~\cite{fetter71}.
The equation of motion method and projection technique are alternative
approaches to obtain the Green functions for interacting 
electrons~\cite{fulde95,fulde12}.  The renormalization-group (RG) 
approach considers successive transformations
to effective Hamiltonians, leading to the same low-energy eigen 
values~\cite{krishna80,shanker94}.
Numerical methods which directly treat the many-body system on 
computers have also been much developed.  In the exact diagonalization
method (ED), one directly solves the eigen-value problem for a small
cluster with use of the Lanczos method~\cite{dagotto94}.  
In the quantum Monte-Carlo
method (QMC), we reduce the quantum mechanical average into a classical
one with use of the Suzuki-Trotter theorem, and apply the Monte-Carlo
technique~\cite{hirsch86,grober00}.  

In the past two decades, effective medium approaches such as
the dynamical coherent potential approximation 
(CPA)~\cite{kake92,kake02} and the dynamical
mean field theory (DMFT)~\cite{georges96,anisimov10} 
have been developed.  
There we replace the surrounding
interactions with a momentum-independent 
effective medium in the Green function and
determine the medium self-consistently solving an impurity problem with
use of many-body techniques.

Among various methods to treat interacting electrons, the variational
wavefunction method is the simplest and oldest 
method~\cite{fulde95,fulde12,fulde02}.  We assume there
a trial wavefunction consisting of the minimum basis set with
variational parameters and determine the parameters on the basis of the
variational principle at the ground-state.
The wavefunction method has a merit of efficiency to best determine the
wavefunction, though one has to find a physically suitable trial
wavefunction or the minimum basis set for correlated electrons.
It also allows us to calculate any static quantities such as the double
occupation number and correlation functions.

The Gutzwiller wavefunction (GW)~\cite{gutz63,gutz64,gutz65} 
is the first which described the
correlated electrons in a narrow band.
In the GW, one varies the amplitudes of doubly-occupied sites which
appear in the Hartree-Fock wavefunction in order to reduce the loss of
Coulomb interaction energy.  The local ansatz (LA)
wavefunction~\cite{stoll77,stoll78,stoll80,fulde95,fulde12} 
is an approach from the weakly correlated limit.  It makes
use of the Hilbert space expanded by the two-particle operators which
appear in the residual Coulomb interactions.
The Baeriswyl wavefunction~\cite{baer87,baer00,hetenyi10} 
expands the Hilbert space with use of the
hopping operators onto the atomic wavefunctions aiming at an accurate
description of electron correlations in the strong interaction regime.
There are various trial wavefunctions which describe the nonlocal electron
correlations~\cite{jast55,kaplan82,yokoyama90,tahara08}.  
These wavefunctions are usually treated by means of
numerical techniques such as the variational Monte-Carlo 
method~\cite{ceperley77,yokoyama87}.

Most of the wavefunctions mentioned above, however, 
do not reduce to the exact
result of the Rayleigh-Schr\"{o}dinger perturbation theory in the weak
Coulomb interaction limit, because they are designed mainly to describe
the electrons in the strong Coulomb interaction regime.
It is indispensable for a quantitative description to construct the
wavefunction leading to the exact result in the weak Coulomb
interaction limit.  In particular, the Fermi liquid state of the 
strongly correlated electrons should be connected 
to that of the weakly correlated
system according to the adiabatic principle for interacting 
electrons~\cite{anderson84},
and thus associated Fermi liquid parameters such as the quasiparticle
weight should be obtained by a suitable renormalization of their 
parameters in the weakly correlated interaction limit.

In this brief review article, we elucidate the momentum-dependent local
ansatz wavefunction (MLA)~\cite{kake08,pat11,pat13,pat13-2} 
which reproduces exactly the weak Coulomb
interaction limit and interpolates between the weak and strong Coulomb
interaction limits.
The MLA is an extension of the LA.  In the MLA, we first expand the
Hilbert space by means of the two-particle excited operators in the
momentum representation, introduce momentum-dependent variational
parameters as the amplitudes of the excited states, and project 
these states
onto the local orbitals again. In this way, we can obtain more flexible
correlated electron states as compared with the LA.  Furthermore we can
improve the MLA by introducing a hybrid (HB) wavefunction as a starting
wavefunction, so that the MLA improves upon both the GW and the LA in
infinite dimensions.  The MLA wavefunction is a counterpart of the DMFT
because it leads to an accurate description of the
Fermi liquid state in infinite dimensions.

In the following section, we briefly review the wavefunction method and
various wavefunctions including the Gutzwiller and LA
wavefunctions, as well as their results for calculations 
of physical quantities.
In Sec. 3, we introduce the MLA wavefunction.  We argue the validity
on the basis of the numerical results in infinite dimensions.  The MLA
does describe well the weak and intermediate Coulomb interaction regimes,
but it does not explain the insulating state in the strong Coulomb
interaction regime.  In Sec. 4, we present the MLA with hybrid (HB)
wavefunction, which allows us to describe both the metallic and
insulator states.  
The idea is to start from the best wavefunction of a hybrid Hamiltonian
obtained by superposing the Hartree-Fock (HF) and the alloy-analogy 
(AA) Hamiltinians.  
In the strong Coulomb interaction region, the AA wavefunction is 
a good starting wavefunction, while
the HF one is the best in the weakly correlated regime.
Since the first-principles GW method does not describe correctly the weak
interaction regime, one needs an alternative 
first-principles wavefunction method which describes quantitatively the
correlated electrons in the weak and intermediate interaction regimes. 
We present an extension of the MLA to the realistic system in 
Sec. 5.  Finally, we summalize the MLA and discuss future problems 
in Sec. 6.

\section{Wavefunction Method and Various Wavefunctions}

The ground-state properties of a system are completely determined by the
wavefunction.  The wavefunction method is based on the variational
principle for the wavefunction.  It states that the expectation value
$E$ of the Hamiltonian $H$ for any trial wavefunction $| \Psi \rangle$
is equal to or larger than the ground-state energy $E_{0}$.
\begin{equation}
E_0 \leq E =\frac{ \langle \Psi |H| \Psi \rangle }
{ \langle \Psi | \Psi \rangle } \, . 
\label{varprin}
\end{equation}
The variational principle allows us to find an approximate but best
wavefunction for a given ansatz, and its energy expectation value
gives us the upper limit of the exact ground-state energy.
In the actual application of Eq. (\ref{varprin}), it is important 
that we adopt a size-consistent wavefunction close to 
the exact one and calculate the energy expectation value 
as accurate as we can in order to avoid uncertainty. 

We consider hereafter the tight-binding model Hamiltonian with
intra-atomic Coulomb interaction called the Hubbard 
model~\cite{hub63,hub65}, for simplicity. 
\begin{eqnarray}
H = \sum_{i \sigma} \epsilon_{0} n_{i\sigma} 
+ \sum_{ij \sigma} t_{i j} \, a_{i \sigma}^{\dagger} a_{j \sigma} 
+ U \sum_{i} \, n_{i \uparrow} n_{i \downarrow} \ .
\label{hub}
\end{eqnarray}
Here $\epsilon_{0}$ is the atomic level, 
$t_{ij}$ is the transfer integral between sites $i$ and $j$.  $U$
is the intra-atomic Coulomb energy parameter.  $ a_{i \sigma}^{\dagger}$
($ a_{i \sigma}$) denotes the creation (annihilation) operator for an
electron on site $i$ with spin $\sigma$, and 
$n_{i\sigma}=a_{i\sigma}^{\dagger} a_{i \sigma}$ denotes the electron density
operator on site $i$ for spin $\sigma$.

The Hamiltonian can be separated into the Hartree-Fock mean-field
Hamiltonian $H_{0}$ and the residual interaction part as follows.
\begin{eqnarray}
H = H_{0} + U \sum_{i} \, O_{i} \ ,
\label{hub2}
\end{eqnarray}
\begin{eqnarray}
H_{0} =  \sum_{i \sigma} ( \epsilon_{0} + U \langle n_{i -\sigma}
 \rangle_{0} ) \, \hat{n}_{i\sigma} +
\sum_{ij \sigma} t_{i j\sigma} \, a_{i \sigma}^{\dagger} a_{j \sigma} 
- U \sum_{i} \, \langle n_{i \uparrow} \rangle_{0} 
\langle n_{i\downarrow} \rangle_{0} \ ,
\label{hf}
\end{eqnarray}
Here $\langle \sim \rangle_{0}$ denotes the Hartree-Fock average at the
ground state. The operator $O_{i}$ in the residual interaction is
defined by  
$O_{i}=\delta n_{i \uparrow}\delta n_{i \downarrow} $ and 
$\delta n_{i\sigma} = n_{i\sigma} - \langle n_{i\sigma} \rangle_{0}$.

The Hartree-Fock ground-state wavefunction $|\phi\rangle$ is given by
\begin{eqnarray}
|\phi\rangle = \Big[ \prod_{k\sigma}^{\rm occ} a^{\dagger}_{k\sigma}
 \Big] |0 \rangle \, .
\label{hfwf}
\end{eqnarray}
Here $\prod_{k\sigma}^{\rm occ}$ means taking the products over the
momentum $k$ and spin $\sigma$ of electrons below the Fermi level.  
$|0 \rangle$
denotes the vacuum state.  $a^{\dagger}_{k\sigma}$ is the creation
operator for an electron with momentum $k$ and spin $\sigma$; 
$a^{\dagger}_{k\sigma} = \sum_{i} a^{\dagger}_{i\sigma} \langle i | 
k \rangle$.  
$\langle i | k \rangle 
( = \exp (-i\boldsymbol{k}\cdot\boldsymbol{R}_{i}) / \sqrt{N} )$ 
is an overlap integral between the localized orbital on site
$i$ and the Bloch state $k$. $\boldsymbol{R}_{i}$ denotes atomic
position of site $i$, and $N$ is the number of sites.

The energy difference between the ground state for the correlated
electrons and the Hartree-Fock one is given by
\begin{eqnarray}
E_{\rm c} = \langle H \rangle -  \langle H \rangle_{0} = 
\dfrac{\langle \Psi |\tilde{H}| \Psi \rangle}{\langle \Psi | \Psi \rangle} \ .
\label{corr}
\end{eqnarray}
Here $\tilde{H}=H - \langle H \rangle_{0}$.
It is the energy gain due to correlated motion of electrons, and is
called the correlation energy.

\subsection{Gutzwiller wavefunction}

The Hartree-Fock mean-field wavefunction overestimates the ground-state
energy because independent motion of electrons produces excessively 
many doubly-occupied
sites with loss of energy due to Coulomb repulsion.
The wavefunction proposed by Gutzwiller~\cite{gutz63,gutz64,gutz65} 
reduces the amplitudes of
doubly occupied states in the Hartree-Fock ground state. 
It is given by
\begin{equation}
|\Psi_{\rm G}\rangle =\Big[\prod_{i}(1-(1-g) n_{i\uparrow}
 {n_{i\downarrow}})\Big]|\phi \rangle \ .
\label{gw}
\end{equation}
The projection operator $n_{i\uparrow} n_{i\downarrow}$ picks 
up the doubly occupied state on site $i$.
The parameter $1-g$ denotes the amplitude of the doubly occupied
states. 
The variational parameter $g=1$ corresponds to the the Hartree-Fock
state, while $g=0$ corresponds to the atomic state with no doubly
occupied state.  Varying the variational parameter $g$ from 1 to 0, one
can choose the best amplitude of doubly occupied states for correlated
electrons on the basis of the variational principle (\ref{varprin}).  

Gutzwiller obtained approximately the ground-state energy by making use
of a quasichemical method~\cite{gutz65}.  The Gutzwiller approximation
was proved to be exact in infinite dimensions~\cite{metzner89}.
In the nonmagnetic state at half-filling, 
we obtain a simple result for the ground-state energy per
atom in infinite dimensions as~\cite{gutz65,kake12}
\begin{eqnarray}
\epsilon_{\rm G}=-\dfrac{1}{8} U_c\left( 1-\dfrac{U}{U_c} \right)^2  \, .
\label{gehalf}
\end{eqnarray}
Here we assumed $\epsilon_{0}=0$.  
$U_{c}=16|\int^{0}_{-\infty} \epsilon \rho(\epsilon) d\epsilon|$
and $\rho(\epsilon)$ is the noninteracting density of states per atom
per spin.  
The ground-state energy increases with increasing $U$ and becomes zero
at $U=U_{c}$.  For $U > U_{c}$, we have a solution $\epsilon_{\rm G}$
with $g=0$.  Therefore the metal-insulator transition occurs at
$U=U_{c}$.  

Similarly the double occupation number per atom linearly decreases
with increasing $U$ at half-filling as
\begin{eqnarray}
d_{\rm G} = \langle n_{i\uparrow} n_{i\downarrow} \rangle 
= \frac{1}{4}\Big(1-\frac{U}{U_c}\Big)\ ,
\label{gdbl}
\end{eqnarray}
and $d_{\rm G}=0$ beyond $U_{c}$.  We call the state $d_{\rm G}=0$ 
the Brinkman-Rice atomic state~\cite{br70}.  It is
therefore realized for $U > U_{c}$, {\it i.e.}, 
in the insulating state.  The Brinkman-Rice atomic state is not
consistent with the super-exchange state with charge fluctuations in the
strongly correlated region. 

The momentum distribution for the
GW is known to be flat below and above the Fermi level, and shows 
a jump at the Fermi level~\cite{gutz65}. 
The latter is the quasiparticle weight 
according to the Fermi liquid theory.  
For half-filling it is given by
\begin{eqnarray}
Z_{\rm G}=1-\frac{U^2}{U^2_c} \, .
\label{gqpw}
\end{eqnarray}
Beyond $U_{c}$, the jump disappears and the distribution becomes
completely flat.  The flat behavior in infinite dimensions is not 
consistent with the result of
the second-order perturbation theory for the Green 
function~\cite{kake04}.

\subsection{Local-ansatz wavefunction}

The Hartree-Fock Hamiltonian neglects the charge (or spin) fluctuations
$\{O_i\}=\{\delta n_{i\downarrow}\delta n_{i\uparrow}\}$ which appear in
the residual interactions.  An alternative way to take into account
electron correlations is therefore to include the Hilbert space expanded
by the fluctuations $\{O_i\}$.  Such a wavefunction is called the local
ansatz (LA)~\cite{stoll77,stoll78,stoll80}.  It is given by
\begin{eqnarray}
|\Psi_{\rm LA}\rangle = \Big[ \prod_{i} (1 - \eta^{}_{\rm \, LA} O_{i})
		    \Big]|\phi \rangle \ .
\label{lawf}
\end{eqnarray}
Here  $\eta^{}_{\rm \, LA}$ is a variational parameter.

In the single-site approximation, the correlation energy per atom is
given as follows~\cite{stoll81,kake88}.
\begin{eqnarray}
\epsilon_{c}({\rm LA}) 
= \dfrac{-2\eta_{\rm LA}\langle O_{i}\tilde{H}\rangle_{0}
+\eta^{2}_{\rm LA}\langle O_{i}\tilde{H}O_{i}\rangle_0}
{1+\eta^{2}_{\rm LA}\langle O_{i}^{2}\rangle_{0}} \ .
\label{lacorr}
\end{eqnarray}
Each element of $\langle O_{i}\tilde{H}\rangle_{0}$, 
$\langle O_{i}\tilde{H}O_{i}\rangle_0$, and 
$\langle O_{i}^{2}\rangle_{0}$ are expressed by the electron number 
$\langle n_{i\sigma} \rangle_{0}$ and the Hartree-Fock local 
density of states $\rho_{i\sigma}(\epsilon)$.
Minimizing the energy $\epsilon_{c}({\rm LA})$ with respect to 
the variational parameter $\eta^{}_{\rm \, LA}$, we obtain
\begin{eqnarray}
\eta_{\rm LA} = 
\frac{\displaystyle -\langle O_{i}\tilde{H}O_{i}\rangle_{0} + 
\sqrt{\langle O_{i}\tilde{H}O_{i}\rangle_{0}^{2}+
4\langle O_{i}\tilde{H}\rangle^{2}_{0}\langle O_{i}^{2}\rangle_{0}}
}
{2\langle O_{i}\tilde{H}\rangle_{0}\langle O_{i}^{2}\rangle_{0}
} \ .
\label{laeta}
\end{eqnarray}

In the nonmagnetic state at half-filling, the double occupation number
in the LA has a simple form,
\begin{eqnarray}
\langle n_{i\uparrow}n_{i\downarrow}\rangle_{\rm LA} = 
\dfrac{1}{4}\Big( 1
-\dfrac{{\eta_{\rm LA}}/{2}}{1+{\eta_{\rm LA}^2}/{16}}\Big)\ .
\label{ladbl}
\end{eqnarray}
The momentum distribution function in the LA also shows a flat behavior
as a function of $\epsilon_{k}$ below and above the Fermi level.  
Here $\epsilon_{k}$ is the Fourier transform of $t_{ij}$.  
The quasiparticle weight as the jump in the momentum distribution 
on the Fermi surface is obtained analytically for half-filling as 
follows.
\begin{eqnarray}
Z_{\rm LA} = 1 - \dfrac{\eta^{2}_{\rm LA}/4}{1+\eta^{2}_{\rm LA}/16} \ .
\label{laqpw}
\end{eqnarray}
Therefore the effective mass diverges at 
$\eta^{\ast}_{\rm LA}=\sqrt{16/3}$ in the LA.

Note that the space expanded by $\{ O_{i} \}$ is not sufficient
to describe the atomic states.  In order to describe the strongly
correlated regime, one has to extend the LA as follows.
\begin{eqnarray}
|\Psi_{\rm LA}\rangle = \Big[ \prod_{i} (1 
- \zeta_{\rm LA} \delta n_{i} - \xi_{\rm LA} \delta m_{i} 
- \eta^{}_{\rm \, LA} O_{i}) \Big]|\phi \rangle \ .
\label{lawfex}
\end{eqnarray}
Here $\zeta_{\rm LA}$ and $\xi_{\rm LA}$ are additional 
parameters controlling the
charge and spin fluctuations, respectively.

The LA is suitable for the description of correlated-electron systems
with a weak or intermediate Coulomb interaction strength, while the
Gutzwiller wavefunction is more suitable in the strongly
correlated region.  More details on the LA and the GW and their
applications to various topics are found in the books by 
Fulde~\cite{fulde95} and 
by Fazekas~\cite{fazekas99}, respectively.

\subsection{Other wavefunctions}

There are many other wavefunctions which have been proposed.
Both the Gutzwiller and the LA wavefunctions do not explicitly include
the inter-site correlation operators.  The wavefunction proposed by
Jastrow~\cite{jast55} 
describes the inter-site density-density correlations and has
the form 
\begin{eqnarray}
|\Psi_{\rm J}\rangle = \exp 
\Big( - \sum_{(i,j)} f_{ij} n_{i} n_{j} \Big)|\phi \rangle
\, .
\label{jawf}
\end{eqnarray}
Here $n_{i} = n_{i\uparrow} + n_{i\downarrow}$ is the density
operator on site $i$ and $f_{ij}$ are variational parameters.
Note that the Gutzwiller wavefunction is expressed as
\begin{eqnarray}
|\Psi_{\rm G}\rangle 
= {\rm e}^{ \eta_{\rm G} \sum_{i} n_{i\uparrow} n_{i\downarrow}} 
|\phi \rangle 
\propto {\rm e}^{ \frac{1}{2} \eta_{\rm G} \sum_{i} n_{i} n_{i}}
|\phi \rangle
\, ,
\label{gjawf}
\end{eqnarray}
where the variational parameters $\eta_{\rm G}$ and $g$ are related
through $\eta_{\rm G} = \ln g$.  Therefore the on-site Jastrow
wavefunction is equivalent to the Gutzwiller wavefunction.

A wavefunction being suitable in the strong correlation regime is the
Baeriswyl wavefunction~\cite{baer87,baer00}.  
It is constructed by applying a hopping
operator 
$\hat{T}=-\sum_{ij\sigma} t_{ij} a^{\dagger}_{i\sigma} a_{j\sigma}$
onto the atomic wavefunction $|\Psi_{\infty} \rangle$ as 
\begin{eqnarray}
|\Psi_{\rm B}\rangle 
= {\rm e}^{ -\eta_{\rm B} \hat{T}} |\Psi_{\infty} \rangle \, .
\label{bawf}
\end{eqnarray}
The operator $\exp (-\eta_{\rm B} \hat{T})$ creates the electron 
hopping states from the atomic one and the variational parameter 
$\eta_{\rm B}$ controls the hopping rate to minimize the energy.

In order to describe the doublon (doubly occupied state)-holon (empty
state) bound state, which appears in the super-exchange process in the
strong Coulomb interaction regime, one can consider the 
wavefunction~\cite{yokoyama90} as
\begin{eqnarray}
|\Psi_{\rm dh}\rangle 
= {\rm e}^{ -\alpha \hat{Q}} |\Psi_{\rm G} \rangle \, .
\label{dhwf}
\end{eqnarray}
Here 
$\hat{Q} = \sum_{i} [ \hat{d}_{i} \prod_{\tau} (1 - \hat{h}_{i+\tau}) 
+ \hat{h}_{i} \prod_{\tau} (1 - \hat{d}_{i+\tau}) ]$. 
$\hat{d}_{i}=n_{i\uparrow} n_{i\downarrow}$ 
($\hat{h}_{i}=(1-n_{i\uparrow})(1-n_{i\downarrow})$) 
is the doublon (holon)
operator, and $\tau$ is taken over the nearest-neighbor sites.
The variational parameter $\alpha$ controls the amplitudes of the
nearest-neighbor doublon-holon bound states.
The ground-state of the non-local wavefunctions are usually calculated
by means of the numerical technique called the variational Monte-Carlo
method (VMC)~\cite{ceperley77,yokoyama87}.

\section{Momentum-Dependent Local Ansatz Wavefunction}

Most of the wavefunctions mentioned in the last section aim 
to describe correlated electrons in the intermediate 
and strong Coulomb interaction regimes.
The behavior of these wavefunctions in the weak Coulomb 
interaction regime was not discussed seriously.
Kakehashi {\it et al.}~\cite{kake08} have 
recently pointed out that the wavefunctions mentioned
above do not yield the exact results in the weak Coulomb interaction
limit according to the Rayleigh-Schr\"{o}dinger perturbation theory of 
the wavefunction.  They proposed a new wavefunction called the
momentum-dependent local ansatz (MLA) which is consistent with the
perturbation theory.  The MLA is a new wavefunction which reproduces 
well-known results in infinite dimensions~\cite{georges96}.  

In the following subsection, we introduce the MLA that describes exactly
the correlated electrons in the weak Coulomb interaction limit, and
elucidate the results obtained by the MLA wavefunction in infinite
dimensions. 

\subsection{Momentum-dependent local ansatz based on the Hartree-Fock 
wavefunction}

The momentum-dependent local ansatz wavefunction (MLA) is constructed
from the local-ansatz (LA) wavefunction (\ref{lawf}) so as to reproduce
the result of the Rayleigh-Schr\"{o}dinger perturbation theory.  We
expand the LA wavefunction (\ref{lawf}) in the weak Coulomb interaction
limit as
\begin{eqnarray}
|\Psi_{\rm LA}\rangle = |\phi \rangle + |\phi_{1}\rangle_{\rm \, LA} 
+ \cdots \ ,
\label{laexp}
\end{eqnarray}
\begin{eqnarray}
|\phi_{1}\rangle_{\rm LA} = 
- \sum_{i} \sum_{k_{1}k^{\prime}_{1}k_{2}k^{\prime}_{2}}
\langle k^{\prime}_{1}|i \rangle \langle i|k_{1} \rangle 
\langle k^{\prime}_{2}|i \rangle \langle i|k_{2} \rangle
\, \eta^{}_{\rm \, LA} \, 
\delta(a^{\dagger}_{k^{\prime}_{2}\downarrow}a_{k_{2}\downarrow})
\delta(a^{\dagger}_{k^{\prime}_{1}\uparrow}a_{k_{1}\uparrow})
|\phi \rangle \, . \ \ 
\label{phi1la}
\end{eqnarray}
Here $\langle i|k \rangle = \exp (-i\boldsymbol{k} \cdot 
\boldsymbol{R}_{i}) / \sqrt{N}$ is an overlap integral 
between the localized orbital on site $i$ and the Bloch state with
momentum $\boldsymbol{k}$, and 
$\delta(a^{\dagger}_{k^{\prime}\sigma}a_{k\sigma})= 
a^{\dagger}_{k^{\prime}\sigma}a_{k\sigma} - \langle
a^{\dagger}_{k^{\prime}\sigma}a_{k\sigma} \rangle_{0}$.

The Rayleigh-Schr\"odinger perturbation theory for the exact
ground-state wavefunction, on the other hand,
yields the following form
\begin{eqnarray}
|\Psi \rangle = |\phi \rangle + |\phi_{1}\rangle + \cdots \ ,
\label{rsexp}
\end{eqnarray}
\begin{eqnarray}
|\phi_{1}\rangle = 
- \sum_{i} \sum_{k_{1}k^{\prime}_{1}k_{2}k^{\prime}_{2}} \!\!\!
\langle k^{\prime}_{1}|i \rangle \langle i|k_{1} \rangle 
\langle k^{\prime}_{2}|i \rangle \langle i|k_{2} \rangle
\, \eta^{(0)}_{k^{\prime}_{2}k_{2}k^{\prime}_{1}k_{1}} 
\delta(a^{\dagger}_{k^{\prime}_{2}\downarrow}a_{k_{2}\downarrow})
\delta(a^{\dagger}_{k^{\prime}_{1}\uparrow}a_{k_{1}\uparrow})
|\phi \rangle \, . \ \ 
\label{phi1}
\end{eqnarray}
The amplitude $\eta^{(0)}_{k^{\prime}_{2}k_{2}k^{\prime}_{1}k_{1}}$ is
given by
\begin{eqnarray}
\eta^{(0)}_{k^{\prime}_{2}k_{2}k^{\prime}_{1}k_{1}} = 
-U \lim_{z \rightarrow 0}
\dfrac{f(\tilde{\epsilon}_{k_{1\uparrow}})
(1-f(\tilde{\epsilon}_{k^{\prime}_{1\uparrow}}))
f(\tilde{\epsilon}_{k_{2\downarrow}})
(1-f(\tilde{\epsilon}_{k^{\prime}_{2\downarrow}}))}
{z - \epsilon_{k^{\prime}_{1\uparrow}} + \epsilon_{k_{1\uparrow}}
- \epsilon_{k^{\prime}_{2\downarrow}} + \epsilon_{k_{2\downarrow}}
} \ .
\label{rseta}
\end{eqnarray}
Here $f(\epsilon)$ is the Fermi distribution function at zero
temperature, and $\tilde{\epsilon}_{k\sigma}=\epsilon_{k\sigma}-\mu$.
$\mu$ is the Fermi level.  $\epsilon_{k\sigma}$ is the
Hartree-Fock one-electron energy eigen value given by 
$\epsilon_{k\sigma}=\epsilon_{0}+U\langle n_{i -\sigma} \rangle_{0} +
\epsilon_{k}$, $\epsilon_{k}$ being the Fourier transform of 
$t_{ij}$.

Comparing Eq. (\ref{phi1}) with Eq. (\ref{phi1la}) indicates that
one has to take into account the momentum dependence of the variational
parameters in order to improve the LA so as to be consistent with the
perturbation theory in the weak Coulomb interaction limit.

In the MLA, we introduce a new local ansatz operator $\tilde{O}_{i}$
such that
\begin{eqnarray}
\tilde{O}_{i} = \sum_{k_{1}k_{2}k^{\prime}_{1}k^{\prime}_{2}} 
\langle k^{\prime}_{1}|i \rangle \langle i|k_{1} \rangle 
\langle k^{\prime}_{2}|i \rangle \langle i|k_{2} \rangle
\eta_{k^{\prime}_{2}k_{2}k^{\prime}_{1}k_{1}} 
\delta(a^{\dagger}_{k^{\prime}_{2}\downarrow}a_{k_{2}\downarrow})
\delta(a^{\dagger}_{k^{\prime}_{1}\uparrow}a_{k_{1}\uparrow}) \ ,
\label{otilde}
\end{eqnarray}
and construct a new wavefunction with 
momentum-dependent variational parameters 
$\{ \eta_{k^{\prime}_{2}k_{2}k^{\prime}_{1}k_{1}} \}$ as 
follows~\cite{kake08}.
\begin{eqnarray}
|\Psi\rangle = \Big[ \prod_{i} (1 - \tilde{O}_{i}) \Big] 
|\phi \rangle \ .
\label{mla}
\end{eqnarray}
The operator $\tilde{O}_{i}$ is still localized on site $i$
because of the projection $\langle k^{\prime}_{1}|i \rangle 
\langle i|k_{1} \rangle \langle k^{\prime}_{2}|i \rangle 
\langle i|k_{2} \rangle$.  
Note that  
$\tilde{O}^{\dagger}_{i} \ne \tilde{O}_{i}$ and 
$\tilde{O}_{i}\tilde{O}_{j} \ne \tilde{O}_{j}\tilde{O}_{i}$
($i \ne j$) in general. These properties however do not cause any
problem when we make a single-site approximation.  
In order that we treat the nonlocal correlations, we have to adopt 
symmetrized operators in general.
Needless to say, the wavefunction $|\Psi\rangle$ reduces to 
the LA $|\Psi_{\rm LA} \rangle$ when the variational 
parameters  $\{ \eta_{k^{\prime}_{2}k_{2}k^{\prime}_{1}k_{1}} \}$ 
are taken to be momentum-independent.

The energy expectation values for the MLA wavefunction can be 
obtained analytically within
the single-site approximation~\cite{kake08}.
Let us consider the numerator 
$A_{N} = \langle \Psi | \tilde{H} | \Psi \rangle$ 
and the denominator $B_{N}$ in the
correlation energy $\langle \tilde{H} \rangle (=A_{N}/B_{N})$. 
\begin{eqnarray}
A_{N} = \Big\langle \Big[\prod_{i} (1-\tilde{O}^{\dagger}_{i}) \Big] 
\tilde{H}
\Big[ \prod_{i} (1-\tilde{O}_{i}) \Big] \Big\rangle_{0}
\ ,
\label{an}
\end{eqnarray}
\begin{eqnarray}
B_{N} = \Big\langle \Big[\prod_{i} (1-\tilde{O}^{\dagger}_{i}) \Big]
\Big[ \prod_{i} (1-\tilde{O}_{i}) \Big] \Big\rangle_{0}
\ .
\label{bn}
\end{eqnarray}
Expanding $B_{N}$ with respect to site 1, we obtain
\begin{eqnarray}
B_{N} & = & B^{(1)}_{N-1} 
- \Big\langle \tilde{O}^{\dagger}_{1}
\Big[{\prod_{i}}^{(1)} (1-\tilde{O}^{\dagger}_{i}) \Big]
\Big[ {\prod_{i}}^{(1)} (1-\tilde{O}_{i}) \Big] \Big\rangle_{0}
\nonumber \\
& &  \hspace{10mm}
- \Big\langle
\Big[{\prod_{i}}^{(1)} (1-\tilde{O}^{\dagger}_{i}) \Big] \tilde{O}_{1}
\Big[ {\prod_{i}}^{(1)} (1-\tilde{O}_{i}) \Big] \Big\rangle_{0}
\nonumber \\
& &  \hspace{10mm}
+ \Big\langle \tilde{O}^{\dagger}_{1}
\Big[{\prod_{i}}^{(1)} (1-\tilde{O}^{\dagger}_{i}) \Big] \tilde{O}_{1}
\Big[ {\prod_{i}}^{(1)} (1-\tilde{O}_{i}) \Big] \Big\rangle_{0}
\ ,  \hspace{10mm}
\label{bn1}
\end{eqnarray}
and
\begin{eqnarray}
B^{(1)}_{N-1} = \Big\langle \Big[ {\prod_{i}}^{(1)} 
(1-\tilde{O}^{\dagger}_{i}) \Big]
\Big[ {\prod_{i}}^{(1)} (1-\tilde{O}_{i}) \Big] \Big\rangle_{0} 
\ .
\label{bn-1}
\end{eqnarray}
Here the product ${\prod_{i}}^{(1)}$ means the products with respect to
all the sites except site 1.

When we apply Wick's theorem for the calculation of $B_{N}$, 
we neglect the contractions between
different sites.  This is a single-site approximation and then
Eq. (\ref{bn1}) is expressed as
\begin{eqnarray}
B_{N} & = & \big\langle 
\big( 1 - \tilde{O}^{\dagger}_{1} \big) 
\big( 1 - \tilde{O}_{1} \big) \big\rangle_{0} \, 
B^{(1)}_{N-1}
\ .
\label{bnssa}
\end{eqnarray}

In the case of $A_{N}$, we take into account the 
two types of terms after application of Wick's theorem, 
the terms in which the operator $\tilde{O}_{1}$ is
contracted to $\tilde{H}$ and the other terms with 
$\tilde{O}_{1}$ contracted to the operators 
$\tilde{O}_{i}$ $(i \ne 1)$.  We have then in the single-site
approximation  
\begin{eqnarray}
A_{N} & = & \big\langle 
\big( 1 - \tilde{O}^{\dagger}_{1} \big) \tilde{H}  
\big( 1 - \tilde{O}_{1} \big) \big\rangle_{0} \, 
B^{(1)}_{N-1}
+ \big\langle 
\big( 1 - \tilde{O}^{\dagger}_{1} \big) 
\big( 1 - \tilde{O}_{1} \big) \big\rangle_{0} \, 
A^{(1)}_{N-1}
\ .
\label{an1}
\end{eqnarray}
Here $A^{(1)}_{N-1}$ is defined by $A^{(1)}_{N}$ in which the correlator
$(1-\tilde{O}^{\dagger}_{1})(1-\tilde{O}_{1})$ on site 1 has been 
removed.

Successive application of the recursive relations (\ref{bnssa}) and
(\ref{an1}) yields  
\begin{eqnarray}
A_{N} & = & \sum_{i} \big\langle 
\big( 1 - \tilde{O}^{\dagger}_{i} \big) \tilde{H}  
\big( 1 - \tilde{O}_{i} \big) \big\rangle_{0} \, 
B^{(i)}_{N-1}
\ ,
\label{an2}
\end{eqnarray}
\begin{eqnarray}
B_{N} 
= \prod_{i} \big\langle 
\big( 1 - \tilde{O}^{\dagger}_{i} \big)  
\big( 1 - \tilde{O}_{i} \big) \big\rangle_{0}
= \big\langle 
\big( 1 - \tilde{O}^{\dagger}_{i} \big)  
\big( 1 - \tilde{O}_{i} \big) \big\rangle_{0} \, 
B^{(i)}_{N-1} 
\ .
\label{bn2}
\end{eqnarray}
Thus we obtain the expression for the correlation energy 
$E_{c}$ as follows.
\begin{eqnarray}
E_{c} = \langle \tilde{H} \rangle =  \sum_{i} \dfrac{
\big\langle 
\big( 1 - \tilde{O}^{\dagger}_{i} \big) \tilde{H}  
\big( 1 - \tilde{O}_{i} \big) \big\rangle_{0}
}
{
\big\langle \big( 1 - \tilde{O}^{\dagger}_{i} \big)  
\big( 1 - \tilde{O}_{i} \big) \big\rangle_{0}
}
\ .
\label{avassa}
\end{eqnarray}
Assuming a site per unit cell and using the relation 
$\langle \tilde{O}^{\dagger}_{i} \rangle_{0} = 
\langle \tilde{O}_{i} \rangle_{0} = 0$, 
we obtain the correlation energy per site as follows.
\begin{eqnarray}
\epsilon_{\rm c} = \dfrac{-\langle
 \tilde{O}^{\dagger}_{i}\tilde{H}\rangle_{0} -
\langle \tilde{H} \tilde{O}_{i} \rangle_{0} + 
\langle \tilde{O}^{\dagger}_{i}\tilde{H}\tilde{O}_{i}\rangle_{0}}
{1 + \langle \tilde{O}^{\dagger}_{i}\tilde{O}_{i} \rangle_{0}} \ .
\label{mlaec}
\end{eqnarray}

By making use of Wick's theorem and the $R=0$ 
approximation~\cite{kajz78,treglia80}, we obtain 
$\langle \tilde{H} \tilde{O}_{i} \rangle_{0} \,( = \langle
\tilde{O}^{\dagger}_{i}\tilde{H} \rangle_{0}^{\ast})$, 
$\langle \tilde{O}^{\dagger}_{i}\tilde{H}\tilde{O}_{i}\rangle_{0}$, 
and
$\langle \tilde{O}^{\dagger}_{i}\tilde{O}_{i} \rangle_{0}$ 
as follows.
\begin{eqnarray}
\langle \tilde{H} \tilde{O}_{i} \rangle_{0} & = &  \frac{U}{N^{4}} 
\sum_{k_{1}k_{2}k^{\prime}_{1}k^{\prime}_{2}} 
\tilde f_{k^{\prime}_{2}k_{2}k^{\prime}_{1}k_{1}} \,
\eta_{k^{\prime}_{2}k_{2}k^{\prime}_{1}k_{1}} \ ,
\label{hor0}
\end{eqnarray}
\begin{eqnarray}
\langle \tilde{O}^{\dagger}_{i}\tilde{H}\tilde{O}_{i}\rangle_{0} & = & 
\dfrac{1}{N^{4}} \sum_{k_{1}k_{2}k^{\prime}_{1}k^{\prime}_{2}} 
\tilde f_{k^{\prime}_{2}k_{2}k^{\prime}_{1}k_{1}}\,
\eta^{\ast}_{k^{\prime}_{2}k_{2}k^{\prime}_{1}k_{1}}  
\bigg[ 
\Delta E_{k^{\prime}_{2}k_{2}k^{\prime}_{1}k_{1}} \,
\eta_{k^{\prime}_{2}k_{2}k^{\prime}_{1}k_{1}}  \nonumber \\
& & \hspace*{-5mm}
+ \dfrac{U}{N^{2}}
\Big\{
\sum_{k_{3}k_{4}}
f(\tilde{\epsilon}_{k_{3}\uparrow})f(\tilde{\epsilon}_{k_{4}\downarrow}) \,
\eta_{k^{\prime}_{2}k_{4}k^{\prime}_{1}k_{3}} 
- \sum_{k_{3}k^{\prime}_{4}}
f(\tilde{\epsilon}_{k_{3}\uparrow}) 
[1-f(\tilde{\epsilon}_{k^{\prime}_{4}\downarrow})] \,
\eta_{k^{\prime}_{4}k_{2}k^{\prime}_{1}k_{3}}  \nonumber \\
& & \hspace*{-18mm}
- \sum_{k^{\prime}_{3}k_{4}}
[1 - f(\tilde{\epsilon}_{k^{\prime}_{3}\uparrow})]
f(\tilde{\epsilon}_{k_{4}\downarrow}) \,
\eta_{k^{\prime}_{2}k_{4}k^{\prime}_{3}k_{1}} 
+ \sum_{k^{\prime}_{3}k^{\prime}_{4}}
[1 - f(\tilde{\epsilon}_{k^{\prime}_{3}\uparrow})]
[1 - f(\tilde{\epsilon}_{k^{\prime}_{4}\downarrow})] \,
\eta_{k^{\prime}_{4}k_{2}k^{\prime}_{3}k_{1}}
\Big\}
\bigg] \, , \nonumber \\
\label{ohor0}
\end{eqnarray}
\begin{eqnarray}
\langle \tilde{O}^{\dagger}_{i}\tilde{O}_{i} \rangle_{0} & = &
\dfrac{1}{N^{4}} \sum_{k_{1}k_{2}k^{\prime}_{1}k^{\prime}_{2}}
|\eta_{k^{\prime}_{2}k_{2}k^{\prime}_{1}k_{1}}|^{2}
\tilde f_{k^{\prime}_{2}k_{2}k^{\prime}_{1}k_{1}} \ . 
\label{oo}
\end{eqnarray}
Here $\Delta E_{k^{\prime}_{2}k_{2}k^{\prime}_{1}k_{1}} =
\epsilon_{k_{2}^{\prime}\downarrow} -
\epsilon_{k_{2}\downarrow} + 
\epsilon_{k_{1}^{\prime}\uparrow} -
\epsilon_{k_{1}\uparrow}$
is a two-particle excitation energy.
$\tilde f_{k^{\prime}_{2}k_{2}k^{\prime}_{1}k_{1}}$ is a 
Fermi factor of two-particle excitations defined by 
$\tilde f_{k^{\prime}_{2}k_{2}k^{\prime}_{1}k_{1}} = 
f(\tilde{\epsilon}_{k_{1\uparrow}})
(1-f(\tilde{\epsilon}_{k^{\prime}_{1\uparrow}}))
f(\tilde{\epsilon}_{k_{2\downarrow}})
(1-f(\tilde{\epsilon}_{k^{\prime}_{2\downarrow}}))$.

Minimizing the correlation energy (\ref{mlaec}), 
we obtain the self-consistent equations for 
$\{ \eta_{k^{\prime}_{2}k_{2}k^{\prime}_{1}k_{1}} \}$ in the single-site
approximation as follows.
\begin{eqnarray}
(\Delta E_{{k^{\prime}_{2}k_{2}k^{\prime}_{1}k_{1}}} - \epsilon_{\rm c})
\eta_{k^{\prime}_{2}k_{2}k^{\prime}_{1}k_{1}}  \hspace{10mm} \nonumber \\
& &  \hspace{-52mm}
+ \dfrac{U}{N^{2}}
\Big[
\sum_{k_{3}k_{4}}
f(\tilde{\epsilon}_{k_{3}\uparrow})f(\tilde{\epsilon}_{k_{4}\downarrow})
\eta_{k^{\prime}_{2}k_{4}k^{\prime}_{1}k_{3}}
\!\! - \!\! \sum_{k_{3}k^{\prime}_{4}}
f(\tilde{\epsilon}_{k_{3}\uparrow})
(1-f(\tilde{\epsilon}_{k^{\prime}_{4}\downarrow}))
\eta_{k^{\prime}_{4}k_{2}k^{\prime}_{1}k_{3}}  \nonumber \\
& & \hspace*{-52mm}
- \sum_{k^{\prime}_{3}k_{4}}
(1 \!\! - \!\! f(\tilde{\epsilon}_{k^{\prime}_{3}\uparrow}))
f(\tilde{\epsilon}_{k_{4}\downarrow})
\eta_{k^{\prime}_{2}k_{4}k^{\prime}_{3}k_{1}}
\!\! + \!\! \sum_{k^{\prime}_{3}k^{\prime}_{4}}
(1 \!\! - \!\! f(\tilde{\epsilon}_{k^{\prime}_{3}\uparrow}))
(1 \!\! - \!\! f(\tilde{\epsilon}_{k_{4}\downarrow}))
\eta_{k^{\prime}_{4}k_{2}k^{\prime}_{3}k_{1}}
\Big] = U . \ \ \
\label{eqeta}
\end{eqnarray}

It is possible to solve approximately the above equation for 
$\eta_{k^{\prime}_{2}k_{2}k^{\prime}_{1}k_{1}}$ for a given
$\epsilon_{c}$.  We first note that 
$\eta_{k^{\prime}_{2}k_{2}k^{\prime}_{1}k_{1}}$ should vanish in the
weak $U$ limit.  Thus, we can omit the second term at the l.h.s. of
Eq. (\ref{eqeta}) in the weak interaction limit.  
Then we obtain the solution as 
$\eta_{k^{\prime}_{2}k_{2}k^{\prime}_{1}k_{1}} = 
U/\Delta E_{k^{\prime}_{2}k_{2}k^{\prime}_{1}k_{1}}$.
In the atomic limit, on the other hand, we have 
$\Delta E_{k^{\prime}_{2}k_{2}k^{\prime}_{1}k_{1}}=0$,
and find a $k$-independent solution $\eta$.  Therefore we approximate 
$\{ \eta_{k^{\prime}_{2}k_{2}k^{\prime}_{1}k_{1}} \}$ in the second term
with a $k$-independent solution $\eta$, so that we obtain an 
approximate solution which interpolates between the weak and 
strong interaction regimes. 
\begin{eqnarray}
\eta_{k^{\prime}_{2}k_{2}k^{\prime}_{1}k_{1}} 
(\tilde{\eta}, \epsilon_{c}) = 
\dfrac{U\tilde{\eta}}
{\Delta E_{{k^{\prime}_{2}k_{2}k^{\prime}_{1}k_{1}}} - \epsilon_{\rm c}} \ .
\label{etaint}
\end{eqnarray}
Here $\tilde{\eta} = 1 - \eta (1 - 2 \langle n_{i\uparrow} \rangle_{0}) 
(1 - 2 \langle n_{i\downarrow} \rangle_{0})$.

When we adopt the approximate form (\ref{etaint}), we have the 
following inequality.
\begin{equation}
E_0 \leq E( \{ \eta_{k'_2 k_2 k'_1 k_1 }^{\ast}\})
\leq E( \{ \eta_{k'_2 k_2 k'_1 k_1 } (\tilde\eta , \epsilon_c)\}) \ ,
\label{ineq2}
\end{equation}
where $\eta_{k'_2 k_2 k'_1 k_1 }^{\ast}$ is the exact stationary value.
Therefore $\tilde{\eta}$ is again determined from the stationary
condition of the correlation energy $\epsilon_{c}$.
\begin{equation}
\tilde \eta=\dfrac{1}{1+\dfrac{U C}{D}} \ .
\label{etatil}
\end{equation}
Here 
\begin{align}
C &= \frac {1}{N^6}\sum_{ k_1 k'_1 k_2 k'_2} 
\frac {\tilde f_{k^{\prime}_{2}k_{2}k^{\prime}_{1}k_{1}}}
{(\Delta E_{k'_2 k_2 k'_1 k_1}-\epsilon_c)} \nonumber \\
&\times \bigg \{  \sum_{k_3 k_4} 
\frac{f( \tilde{\epsilon}_{k_3 \uparrow}) f( \tilde{\epsilon}_{ k_4 \downarrow} )} {(\Delta E_{k'_2 k_4 k'_1 k_3}-\epsilon_c)}
-\sum_{k'_3 k_4}\frac {[1-f( \tilde{\epsilon}_{k'_3\uparrow})] 
f( \tilde{\epsilon}_{ k_4 \downarrow} )}{(\Delta E_{k'_2 k_4 k'_3 k_1}-\epsilon_c)}\nonumber \\
&-\sum_{k_3 k'_4} \frac{f( \tilde{\epsilon}_{k_3 \uparrow})
[1-f( \tilde{\epsilon}_{ k'_4 \downarrow} )]} {(\Delta E_{k'_4 k_2 k'_1 k_3}-\epsilon_c)} 
+ \sum_{k'_3 k'_4} \frac{[1-f(\tilde{\epsilon}_{k'_3 \uparrow})] [1-f(\tilde{\epsilon}_{{k'_4}\downarrow} ) ]}
{(\Delta E_{k'_4 k_2 k'_3 k_1}-\epsilon_c)} \bigg \},
\label{etatilde}
\end {align} 
\begin{equation}
D = \frac {1}{N^4}\sum_{ k_1 k'_1 k_2 k'_2} 
\frac {\tilde f_{k^{\prime}_{2}k_{2}k^{\prime}_{1}k_{1}}}
{(\Delta E_{k'_2 k_2 k'_1 k_1}-\epsilon_c)} \, .
 \hspace{50mm}
\label{etatilnu}
\end {equation}

Note that $\tilde{\eta}$ in Eq. (\ref{etatil}) is given as a function of
$\epsilon_{c}$, and $\epsilon_{c}$ in Eq. (\ref{mlaec}) depends on
$\tilde{\eta}$ and $\epsilon_{c}$.  Thus both equations are solved
self-consistently.  This is the self-consistent MLA which starts from 
the Hartree-Fock wavefunction.  The self-consistency is significant 
when the average electron number deviates from half-filling.

In the numerical calculations of $C, D, \langle \tilde{H} \tilde{O}_{i}
\rangle_{0},  \langle
\tilde{O}^{\dagger}_{i}\tilde{H}\tilde{O}_{i}\rangle_{0}$, and 
$\langle \tilde{O}^{\dagger}_{i}\tilde{O}_{i} \rangle_{0}$, the six-fold
$k$ sums appear. This means that one has to perform the six-fold
integrals in the energy representation.  One can reduce the six-fold
integrals into two-fold ones by using a Laplace 
transformation~\cite{schweiz91}. 
\begin{eqnarray}
\dfrac{1}{z - \epsilon_{4} + \epsilon_{3} - \epsilon_{2} + \epsilon_{1}
 + \epsilon_{c}} = -i \int^{\infty}_{0} dt \,
{\rm
e}^{i(z-\epsilon_{4}+\epsilon_{3}-\epsilon_{2}+\epsilon_{1}
+\epsilon_{\rm c}) \, t}
\ .
\label{laplace}
\end{eqnarray}
Here $z=\omega+i\delta$, and $\delta$ is an infinitesimal positive
number.

Electron number $\langle n_{i} \rangle$, momentum distribution 
$\langle n_{k\sigma} \rangle$, and double occupation number 
$\langle n_{i\uparrow} n_{i\downarrow} \rangle$ are obtained from 
$\langle H \rangle$ by taking the derivative with respect to 
$\epsilon_{0}$ on site $i$, $\epsilon_{k}-\sigma h$, and $U_{i}$, 
respectively.
Here we added the external magnetic field $h$ in the atomic level, 
and the site index $i$ in the Coulomb energy parameter $U$ 
for convenience.
\begin{equation}
\langle n_i \rangle = \langle n_i \rangle_0+\frac{\sum_\sigma \langle \tilde{O}_i^\dagger \tilde{n}_{i\sigma} \tilde{O_i} \rangle_0
}{1+\langle \tilde{O}_i^\dagger  \tilde{O_i} \rangle_0
} \, , 
\label{nave1}
\end{equation}
\begin{equation}
\langle n_{k\sigma} \rangle =\langle n_{k \sigma} \rangle_0 
+\frac{N\langle \tilde{O}_i^\dagger \tilde{n}_{k\sigma}\tilde{O_i} 
\rangle_0}{1+\langle \tilde{O_i}^\dagger \tilde{O_i} \rangle_0} \, , 
\label{nkave}
\end{equation}
\begin{align}
\langle n_{i\uparrow}n_{i\downarrow} \rangle = 
\langle n_{i\uparrow} \rangle_{0} \langle n_{i\downarrow} \rangle_{0}
+ \langle n_{i\uparrow}n_{i\downarrow} \rangle_{c} \ , 
\label{dblav}
\end{align}
\begin{align}
\langle n_{i\uparrow}n_{i\downarrow} \rangle_{c} =  
\dfrac{-\langle \tilde{O}^{\dagger}_{i} O_{i} \rangle_{0} 
- \langle O_{i} \tilde{O}_{i} \rangle_{0} 
+ \langle \tilde{O}^{\dagger}_{i} O_{i} \tilde{O}_{i} \rangle_{0}
+ \sum_{\sigma} \langle n_{i-\sigma} \rangle_{0} 
\langle \tilde{O}^{\dagger}_{i} \tilde{n}_{i\sigma} 
\tilde{O}_{i} \rangle_{0}
}
{1+\langle \tilde{O}^{\dagger}_{i} \tilde{O}_{i} \rangle_{0}} \ .
\label{dblavc}
\end{align}
The second terms at the r.h.s. of the above expressions 
(\ref{nave1}), (\ref{nkave}), and (\ref{dblav}) 
are correlation
corrections and they are again calculated by using the Laplace
transformation.

\subsection{MLA in infinite dimensions}

The MLA improves upon the LA irrespective of the Coulomb 
interaction strength
and the electron number.  One can demonstrate this fact by means of 
numerical calculations in infinite dimensions. We adopt hereafter the
Hubbard model on the hypercubic lattice in infinite 
dimensions~\cite{muller89}.  In this
case, the density of states (DOS) for the noninteracting system 
is given by
$\rho(\epsilon) = (1/\sqrt{\pi})\exp(-\epsilon^{2})$.  
%
%
\begin{figure}[htbp]
\begin{center}
\includegraphics[scale=0.80]{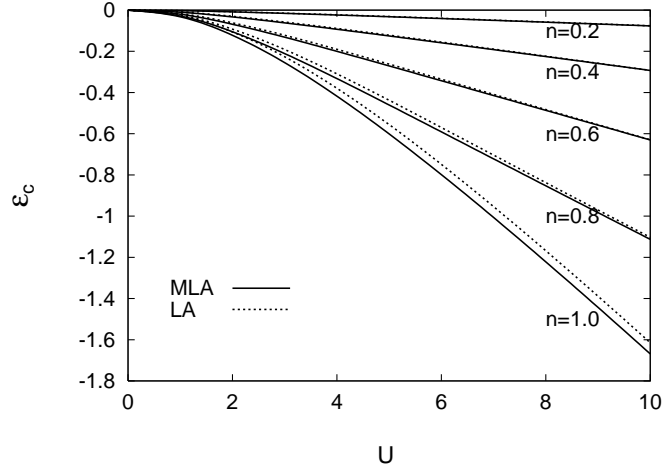} 
\caption{\label{figec}
The correlation energies $\epsilon_{c}$ per atom vs. Coulomb 
interaction energy parameter 
$U$ in the MLA (solid curve) and the LA (dashed curve) for various 
electron number $n$ on the hypercubic lattice in infinite 
dimensions (Ref. 45).
 The energy unit is chosen so that the band width of noninteracting
 system be 2.
}
\end{center}
\end{figure}
%
%

Figure \ref{figec} shows the correlation energy per atom as a 
function of $U$ for various electron numbers. 
We verify that the ground-state energy in the MLA is lower
than that of the LA over all Coulomb interactions $U$ and 
electron numbers $n$.  
In particular, the small $U$ behavior of $\epsilon_{c}$ in the MLA
is exact.  For a given $U$, the difference between the LA and the MLA
increases with increasing $n$ and becomes maximum at half-filling 
because the number of doubly occupied sites in the Hartree-Fock ground
state increases with increasing electron number.

\begin{figure}[htbp]
\begin{center}
\includegraphics [scale=0.80]{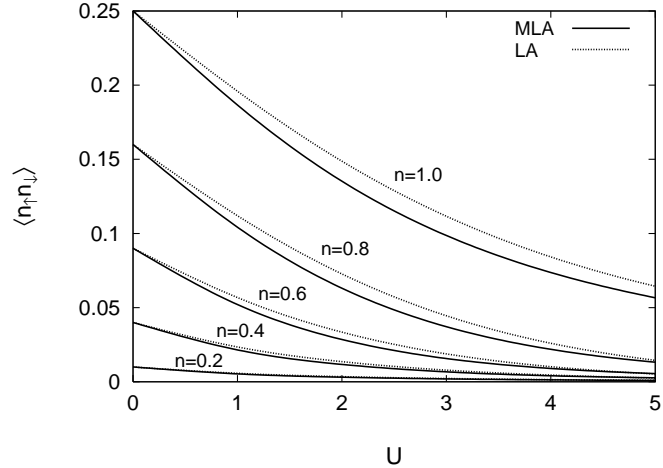}%
\caption{\label{figdbl}
The double occupation number 
$\langle n_{\uparrow}n_{\downarrow} \rangle$ 
vs. Coulomb interaction energy $U$ curves 
in the MLA (solid curve) and the LA (dotted curve) (Ref. 45).}
\end{center}
\end{figure}

The double occupation number decreases with increasing interaction 
$U$ irrespective of electron number $n$ so as to suppress the loss 
of Coulomb interaction energy as seen in Fig. \ref{figdbl}.
We find that the MLA wavefunction gives greater reduction in 
the double occupancy as compared with that of the LA.
%
%
%
%

\begin{figure}[htbp]
\begin{center}
\includegraphics[scale=0.80]{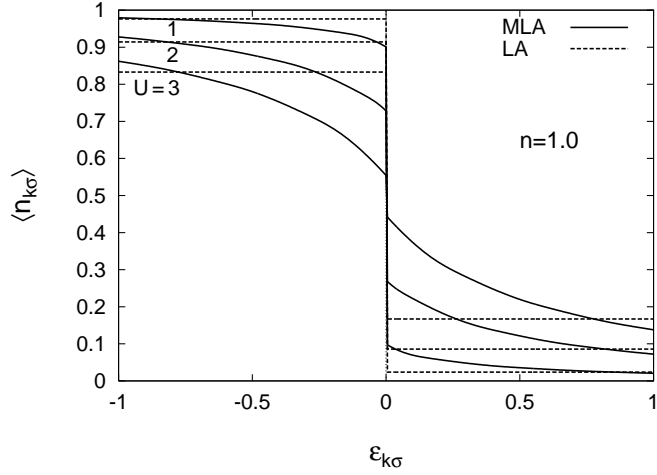} 
\caption{\label{fignk}
The momentum distribution as a function of energy $\epsilon_{k\sigma}$ 
for various Coulomb interaction energy 
parameters $U$ at half-filling ($n=1.0$) (Ref. 45). 
The MLA: solid curves, the LA: dashed curves.}
\end{center}
\end{figure}

The momentum dependence of the variational parameters causes
qualitative change in the momentum distribution as shown in Fig. 
\ref{fignk}.  The momentum distribution in the LA and the GW are
constant below and above the Fermi level as mentioned in the last
section, while the distribution in the MLA monotonically decreases 
with increasing energy $\epsilon_{k\sigma}$ below and above 
the Fermi level, as it should~\cite{kake04}. 
%
%
%
%

\begin{figure}[htbp]
\begin{center}
\includegraphics[scale=0.80]{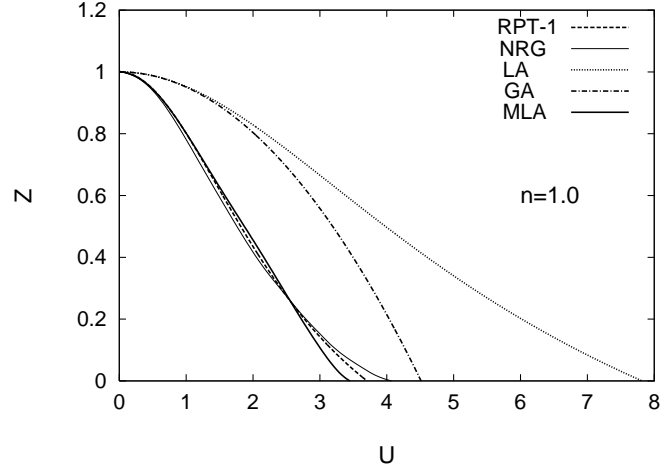}
\caption{\label{figzu1}
Quasiparticle-weight vs. Coulomb interaction curves in various
 theories (Ref. 45). The RPT(Renormalized Perturbation Theory)-1: 
dashed curve (Ref. 52), the NRG: thin solid 
curve (Ref. 60), the LA: dotted curve, the MLA: solid curve, and 
the GW: dot-dashed curve.}
\end{center}
\end{figure}

The quasi-particle weight obtained from the jump in the momentum
distribution at the Fermi level is
also much improved by taking into account the momentum dependence of
variational parameters.
Figure \ref{figzu1} shows the quasi-particle weight $Z$ as a function of
the Coulomb interaction strength $U$ in various
methods at half filling.  
The  quasiparticle weight in the LA changes as
$Z_{\rm LA}=(1-3\eta^{2}_{\rm \, LA}/16)/(1+\eta^{2}_{\rm \, LA}/16)$ 
(see Eq. (\ref{laqpw})) and vanishes at 
$U_{\rm c2}(\rm LA) = 24/\sqrt{3\pi} \, (=7.82)$. 
In the GW~\cite{br70}, the quasiparticle weight decreases as
$Z_{\rm G}=1-(U/U_{\rm c2})^{2}$ (see Eq. (\ref{gqpw})), and 
vanishes at $U_{\rm c2}(\rm GW)=8/\sqrt{\pi} \, (=4.51)$.
These curves deviate strongly from the curve obtained by the numerical
renormalization group method (NRG)~\cite{bulla99} 
which is considered to be the best.
The curve in the MLA on the other hand is close to the that of 
the NRG, and significantly improves upon the LA, 
though calculated $U_{c2}({\rm MLA}) = 3.40$ is
somewhat smaller than the value $U_{c2}({\rm NRG}) = 4.10$.  
%
%
%
%

The numerical results mentioned above indicate that the momentum
dependence of the variational parameters much improves upon the LA as 
well as the GW in the metallic region. In particular, this is 
significant in order to describe the properties associated with the 
low-energy excitations.

\section{Momentum Dependent Local Ansatz with Hybrid Wavefunction}

The MLA describes the electron
correlations in the weak Coulomb interaction limit exactly, and much
improves the LA wavefunction, as we have seen in the last section.
It cannot, however, suppress sufficiently the loss of Coulomb
interaction energy in the strongly correlated region.  The usual way to
take into account more correlations 
is to expand the Hilbert space applying additional correlators with 
variational parameters onto the Hartree-Fock wavefunction.  
In particular, the correlator which suppresses 
the double occupancy is required in the strongly correlated regime.
Such an extention, however, would make it more difficult to treat the
wavefunction analytically.  
An alternative way to overcome the difficulty 
is to start from a wavefunction which is more suitable for
the strongly correlated electrons.  In this section we present an
improvement of the MLA from the latter point of view.

\subsection{Alloy analogy wavefunction}

The Hartree-Fock approximation is exact in energy up to the first
order with respect to the Coulomb interaction energy, therefore 
the wavefunction is
suitable as a starting state for describing correlations in the weak and
intermediate Coulomb interaction regime.  
However, the wavefunction is not
suitable in the strongly correlated region because it allows for the
double occupation of electrons at each site.  

Hubbard proposed an alternative one-electron picture in the strong
Coulomb interaction regime~\cite{hub65}.  
Let us consider the atomic limit.  There
each electron number $\hat{n}_{i\sigma}$ is a good quantum number taking
a value $n_{i\sigma} =$ 0 or 1.  Here and in the following we
distinguish the number operator $\hat{n}_{i\sigma}$
 with the c-number $n_{i\sigma}$(= 0 or 1).  
When the electron hopping is switched on in the strongly
correlated region, an electron with spin $\sigma$ should move slowly
from site to site, and feel a different potential $U n_{i-\sigma} =$ 
$U$ or 0, instead of the Hartree-Fock average potential 
$U\langle \hat{n}_{i-\sigma} \rangle_{0}$, 
depending on whether the opposite-spin electron is occupied or
unoccupied on the same site.
Hubbard regarded the system as an alloy with different random potentials
$\epsilon_{0}+U$ and $\epsilon_{0}$ having the concentration 
$\langle \hat{n}_{i -\sigma} \rangle$ (occupied) and 
$1 - \langle \hat{n}_{i -\sigma} \rangle$ (unoccupied), respectively.
This is the alloy-analogy (AA) picture for strongly correlated
electrons.

The AA Hamiltonian is given by
\begin{eqnarray}
H_{\rm AA} & = & \sum_{i\sigma}(\epsilon_0 +  
U n_{i-\sigma})\hat n_{i\sigma}
+ \sum_{ij\sigma} t_{ij}  a_{i\sigma}^\dagger  a_{j\sigma} 
-U \sum_i (n_{i\uparrow}\langle \hat n_{i \downarrow} \rangle_{\rm AA}
\! + \! n_{i\downarrow} \langle \hat n_{i \uparrow} \rangle_{\rm AA}) 
\nonumber \\
& & \hspace{58mm} + U \sum_i \langle \hat n_{i \uparrow} \rangle_{\rm AA}
\langle \hat n_{i \downarrow} \rangle_{\rm AA}.
\label{aah}
\end{eqnarray} 
Here $\langle \sim \rangle_{\rm AA}$ denotes the AA average 
$\langle \phi_{\rm AA}| (\sim) |\phi_{\rm AA}\rangle$ with respect to
the ground-state wavefunction $|\phi_{\rm AA}\rangle$ of the AA
Hamiltonian $H_{\rm AA}$.  $n_{i\sigma}$ is a c-number taking a value 0
or 1.  Each configuration $\{ n_{i\sigma} \}$ is considered as a
snapshot in time development.

The ground-state energy $E_{0}$ satisfies the following 
inequality for any
configuration of $\{ n_{i\sigma} \}$.
\begin{eqnarray}
E_{0} \leq \langle H \rangle_{\rm AA} = 
\langle H_{\rm AA} \rangle_{\rm AA}\ .
\label{aaineq}
\end{eqnarray}
Thus, when we take the configurational average on 
$\{ n_{i\sigma} \}$, we have  
\begin{eqnarray}
E_{0} \leq \overline{\langle H \rangle}_{\rm AA} \, .
\end{eqnarray}
Here the upper bar denotes the configurational average.

The configurational averages of various quantities can be obtained 
with use of the single-site approximation (SSA) called the
coherent potential approximation (CPA)~\cite{velicky68,shiba71,ehren76}.  
Note that the averaged
electron number is obtained from the local density of state (LDOS) for
an electron with spin $\sigma$, {\it i.e.}$, 
\rho_{i\sigma}(\epsilon)$, as follows.
\begin{eqnarray}
\langle \hat n_{i\sigma}\rangle_{\rm AA} = \int f(\epsilon) 
\rho_{i\sigma}(\epsilon) \, d\epsilon  \ ,
\label{aaavn}
\end{eqnarray} 
and the LDOS is obtained from the one-electron Green function as
\begin{eqnarray}
\rho_{i\sigma}(\epsilon)= - \dfrac{1}{\pi}\, {\rm Im} \, 
G_{ii\sigma}(z) \ .
\label{aaldos}
\end{eqnarray}
The Green function $G_{ii\sigma}(z)$ is defined by 
\begin{eqnarray}
 G_{ii\sigma}(z)= [(z-\boldsymbol{H}_\sigma)^{-1} ]_{ii}\ .
\label{aagii}
\end{eqnarray}
Here $(\boldsymbol{H}_{\sigma})_{ij}$ is the one-electron 
Hamiltonian matrix for the AA Hamiltonian minus chemical potential 
$\mu$.

In the CPA, we replace the random potential at the surrounding sites
with the energy-dependent coherent potential $\Sigma_{\sigma}(z)$.  The
on-site impurity Green function $G_{ii\sigma}(z)$ is then obtained as
follows. 
\begin{eqnarray}
G_{ii\sigma}(z) = \frac{1}{F_{\sigma}(z)^{-1} - \epsilon_{i\sigma} + 
\Sigma_\sigma(z)} \ .
\label{aaimpg}
\end{eqnarray}
Here $\epsilon_{i\sigma}=\epsilon_{0}-\mu+Un_{i -\sigma}$. 
$F_{\sigma}(z)$ is the on-site Green function for the coherent system 
in which all the random potentials have been replaced by the coherent 
potentials.
\begin{eqnarray}
F_{\sigma}(z)= 
\int \frac{\rho(\epsilon) \, d\epsilon}
{z - \Sigma_{\sigma}(z)-\epsilon} \ .
\label{acohg2}
\end{eqnarray}
Note that $\rho(\epsilon)$ is the DOS per site per spin 
for the noninteracting system.  
The coherent potential $\Sigma_{\sigma}(z)$ is determined 
from the self-consistent condition.
\begin{eqnarray}
\overline{G_{00\sigma}(z)} = F_\sigma(z) \ .
\label{acpa}
\end{eqnarray}
The configurational average of  the impurity Green function is 
given as  
\begin{eqnarray}
\overline{G_{00\sigma}(z)}=\frac{\langle \hat n_{i -\sigma} 
\rangle_{\rm AA}}{F_{\sigma}(z)^{-1}-\epsilon_{0}+\mu - U + 
\Sigma_\sigma(z)}
+ \frac{1-\langle \hat n_{i -\sigma}\rangle_{\rm AA}}
{F_{\sigma}(z)^{-1}-\epsilon_{0}+\mu +\Sigma_\sigma(z)}
\label{agf}
\end{eqnarray}

The ground-state wavefunction $\phi_{\rm AA}$ 
for the alloy-analogy Hamiltonian
(\ref{aah}) provides us with a good starting wave function for the
strongly correlated electrons, though such a wavefunction depends 
on electron configuration $\{ n_{i\sigma} \}$ via atomic potentials.

\subsection{MLA with hybrid wavefunction}

We can improve the MLA correlated wavefunction 
using the best starting wavefunction.  
The Hartree-Fock (HF) wavefunction 
$|\phi_{\rm HF}\rangle (=|\phi \rangle)$ works best in 
the weakly correlated region.  In the strongly correlated region the
alloy-analogy (AA) wavefunction $|\phi_{\rm AA}\rangle$ works better.
Therefore we introduce a hybrid (HB) wavefunction $|\phi_{\rm HB}\rangle$
which is the ground state of a hybrid Hamiltonian $H_{\rm HB}$.
The Hamiltonian is defined by a linear combination of the HF and AA
Hamiltonians~\cite{pat13-2}.
\begin{eqnarray}
H_{\rm HB}& = & \sum_{i\sigma}  (\epsilon_{0} 
+ \overline U\langle n_{i-\sigma} \rangle_{\rm HB}
+ \widetilde U n_{i-\sigma}) \hat n_{i\sigma} 
+ \sum_{ij \sigma} t_{i j} \, a_{i \sigma}^{\dagger} 
a_{j \sigma} \nonumber \\
& & \hspace*{0mm} 
- (\overline U- \widetilde U) \sum_{i} \, 
\langle \hat{n}_{i \uparrow} \rangle_{\rm HB} 
\langle \hat{n}_{i\downarrow} \rangle_{\rm HB}
-\widetilde U \sum_i (n_{i\uparrow} 
\langle \hat{n}_{i \downarrow} \rangle_{\rm HB}
+ n_{i\downarrow} \langle \hat{n}_{i \uparrow} \rangle_{\rm HB}) \, .
\label{hhb}
\end{eqnarray}
Here $\langle \sim \rangle_{\rm HB}$ denotes the HB average 
$\langle \phi_{\rm HB}| (\sim) |\phi_{\rm HB}\rangle$, 
$\overline U=(1-w)U$, and $\widetilde U=w U$. $w$ is a weight in 
the linear combination; 
$H_{\rm HB}=(1-w)H_{\rm HF}+w H_{\rm AA} $, where $H_{\rm HF}$ denotes
the HF Hamiltonian. $H_{\rm HB}$ reduces to the  HF (AA)
Hamiltonian when $w=0$ ($w=1$).

The new MLA with the HB wavefunction is given by
\begin {equation}
 | \Psi \rangle = \Big[ \prod_i (1-\tilde{O}_{i}) \Big] 
| \phi_{\rm HB} \rangle .
\label {hwmlahb} 
\end{equation}  
The local operators $\{ \tilde{O}_{i} \}$ have been modified as follows.
\begin{eqnarray}
\tilde{O_i} = \sum_{\kappa'_2 \kappa_2 \kappa'_1 \kappa_1} 
\langle \kappa'_1 | i\rangle \langle i | \kappa_1 \rangle 
\langle \kappa'_2 | i\rangle \langle i | \kappa_2 \rangle 
\ \eta_{\kappa'_2 \kappa_2 \kappa'_1 \kappa_1} 
\delta ({a_{\kappa'_2 \downarrow}^\dagger} a_{\kappa_2 \downarrow }) 
\delta ({a_{\kappa'_1 \uparrow }^\dagger} a_{\kappa_1 \uparrow }) \ . 
\label{hboi}
\end{eqnarray} 
Here $\eta_{\kappa'_2 \kappa_2 \kappa'_1 \kappa_1}$ is a variational 
parameter,
$ a_{\kappa \sigma}^{\dagger}$ and $ a_{\kappa \sigma}$ are 
the creation and annihilation operators which diagonalize 
the Hamiltonian $H_{\rm HB}$ (\ref{hhb}), 
and $\langle \kappa | i \rangle$ are overlap integrals defined by 
$a_{\kappa\sigma} = \sum_{i} a_{i\sigma} \langle \kappa | i \rangle$.
Furthermore  
$\delta(a^{\dagger}_{\kappa^{\prime}\sigma}a_{\kappa\sigma})= 
a^{\dagger}_{\kappa^{\prime}\sigma}a_{\kappa\sigma} - \langle
a^{\dagger}_{\kappa^{\prime}\sigma}a_{\kappa\sigma} \rangle_{\rm HB}$.

The ground-state energy $E_{0}$ again satisfies the following 
inequality for any wavefunction $| \Psi \rangle$:
\begin{eqnarray}
E_{0} \leq \langle H \rangle_{\rm HB}+ N \epsilon_c \ .
\label{hbineq2}
\end{eqnarray}
The correlation energy per atom $\epsilon_{c}$ in the single-site
approximation (SSA) is given as follows (see Eq. (\ref{mlaec})).
\begin{eqnarray}
\epsilon_{\rm c} = \dfrac{-\langle
 \tilde{O}^{\dagger}_{i}\tilde{H}\rangle_{\rm HB} -
\langle \tilde{H} \tilde{O}_{i} \rangle_{\rm HB} + 
\langle \tilde{O}^{\dagger}_{i}\tilde{H}\tilde{O}_{i}\rangle_{\rm HB}}
{1 + \langle \tilde{O}^{\dagger}_{i}\tilde{O}_{i} \rangle_{\rm HB}} \ .
\label{hec}
\end{eqnarray}
Here $\tilde{H} = H - \langle H \rangle_{\rm HB}$.

The energy elements 
$\langle \tilde{H} \tilde{O}_{i} \rangle_{\rm HB}$, 
$\langle \tilde{O}^{\dagger}_{i}\tilde{H}\tilde{O}_{i}\rangle_{\rm HB}$, 
and 
$\langle \tilde{O}^{\dagger}_{i}\tilde{O}_{i} \rangle_{\rm HB}$ 
are given by
\begin{eqnarray}
\langle \tilde{H} \tilde{O}_{i} \rangle_{\rm HB}  = U  \sum_{\kappa'_2 \kappa_2 \kappa'_1 \kappa_1}
|\langle \kappa'_1 | i\rangle|^{2} |\langle \kappa_1 |i  \rangle|^{2}
|\langle \kappa'_2 | i\rangle|^{2} |\langle \kappa_2 |i  \rangle|^{2}\,
\eta_{\kappa'_2 \kappa_2 \kappa'_1 \kappa_1} \,
\tilde f_{\kappa'_2 \kappa_2 \kappa'_1 \kappa_1}  \ ,
\label{hhor0}
\end{eqnarray}
\begin{eqnarray}
\langle \tilde{O}^{\dagger}_{i}\tilde{H}\tilde{O}_{i}\rangle_{\rm HB} & = & 
\sum_{\kappa'_2 \kappa_2 \kappa'_1 \kappa_1}
|\langle \kappa'_1 | i\rangle|^{2} |\langle \kappa_1 |i  \rangle|^{2}
|\langle \kappa'_2 | i\rangle|^{2} |\langle \kappa_2 |i  \rangle|^{2}\,\nonumber \\
& & \hspace*{7mm}
\times
\eta^{\ast}_{\kappa'_2 \kappa_2 \kappa'_1 \kappa_1} \,
\tilde f_{\kappa'_2 \kappa_2 \kappa'_1 \kappa_1} 
\bigg[ 
\Delta E_{\kappa'_{2}\kappa_{2}\kappa'_{1}\kappa_{1}} 
\eta_{\kappa'_2 \kappa_2 \kappa'_1 \kappa_1} \nonumber \\
& & \hspace*{7mm}
+ U \Big\{
\sum_{\kappa_{3}\kappa_{4}}|\langle \kappa_3 | i\rangle|^{2} |\langle \kappa_4 | i \rangle|^{2}
f(\tilde{\epsilon}_{\kappa_{3}\uparrow})f(\tilde{\epsilon}_{\kappa_{4}\downarrow}) \,
\eta_{\kappa'_2 \kappa_4 \kappa'_1 \kappa_3}\nonumber \\
& &\hspace*{7mm}
- \sum_{\kappa'_{3}\kappa_{4}}|\langle \kappa'_3 | i\rangle|^{2} |\langle \kappa_4 | i \rangle|^{2}
(1-f(\tilde{\epsilon}_{\kappa'_{3}\uparrow}))f(\tilde{\epsilon}_{\kappa_{4}\downarrow})\,
\eta_{\kappa'_2 \kappa_4 \kappa'_3 \kappa_1}  \nonumber \\
& & \hspace*{7mm}
- \sum_{\kappa_{3}\kappa'_{4}}|\langle \kappa_3 | i\rangle|^{2} |\langle \kappa'_4 | i \rangle|^{2}
 f(\tilde{\epsilon}_{\kappa_{3}\uparrow})(1-f(\tilde{\epsilon}_{\kappa'_{4}\downarrow}))\,
\eta_{\kappa'_4 \kappa_2 \kappa'_1 \kappa_3}\nonumber \\
& &\hspace*{-7mm}
+ \sum_{\kappa'_{3}\kappa'_{4}}|\langle \kappa'_3 | i\rangle|^{2} |\langle \kappa'_4 | i \rangle|^{2}
 (1-f(\tilde{\epsilon}_{\kappa'_{3}\uparrow}))(1-f(\tilde{\epsilon}_{\kappa'_{4}\downarrow}))\,
\eta_{\kappa'_4 \kappa_2 \kappa'_3 \kappa_1}
\Big\}
\bigg] \ ,
\label{hohor0}
\end{eqnarray}
\begin{eqnarray}
\langle \tilde{O}^{\dagger}_{i}\tilde{O}_{i} \rangle_{\rm HB} & = &
\!\! \sum_{\kappa'_2 \kappa_2 \kappa'_1 \kappa_1} \!\!
|\langle \kappa'_1 | i\rangle|^{2} |\langle \kappa_1 |i  \rangle|^{2}
|\langle \kappa'_2 | i\rangle|^{2} |\langle \kappa_2 |i  \rangle|^{2}\,
|\eta_{\kappa'_2 \kappa_2 \kappa'_1 \kappa_1}|^{2} \,
\tilde f_{\kappa'_2 \kappa_2 \kappa'_1 \kappa_1} \, . 
\hspace{00mm}
\label{hoo}
\end{eqnarray}
Here $\tilde{f}_{\kappa'_2 \kappa_2 \kappa'_1 \kappa_1}$ is the Fermi
factor; 
$\tilde f_{\kappa^{\prime}_{2}\kappa_{2}k^{\prime}_{1}\kappa_{1}} = 
f(\tilde{\epsilon}_{\kappa_{1\uparrow}})
(1-f(\tilde{\epsilon}_{\kappa^{\prime}_{1\uparrow}}))
f(\tilde{\epsilon}_{\kappa_{2\downarrow}})
(1-f(\tilde{\epsilon}_{\kappa^{\prime}_{2\downarrow}}))$.
$\tilde{\epsilon}_{\kappa\sigma}=\epsilon_{\kappa\sigma}-\mu$, 
$\epsilon_{\kappa\sigma}$ being the one-electron energy eigen value 
for $H_{\rm HB}$.
$\Delta 
E_{\kappa^{\prime}_{2}\kappa_{2}\kappa^{\prime}_{1}\kappa_{1}} =
\epsilon_{\kappa_{2}^{\prime}\downarrow} -
\epsilon_{\kappa_{2}\downarrow} + 
\epsilon_{\kappa_{1}^{\prime}\uparrow} -
\epsilon_{\kappa_{1}\uparrow}$
denotes the two-particle excitation energy from the ground state 
$|\phi_{\rm HB} \rangle$.
Note that the above expressions reduce to Eqs. (\ref{hor0}), 
(\ref{ohor0}), and (\ref{oo}), therefore the correlation 
energy (\ref{hec}) reduces to Eq. (\ref{mlaec}) when 
$w \rightarrow 0$.

From the stationary condition $\delta \epsilon_{c} = 0$, we obtain the
self-consistent equations for 
$\{ \eta_{\kappa'_2 \kappa_2 \kappa'_1 \kappa_1}\}$, and again obtain 
an approximate form (see Eq. (\ref{etaint})) such as 
\begin{eqnarray}
\eta_{\kappa^{\prime}_{2}\kappa_{2}\kappa^{\prime}_{1}\kappa_{1}}(\tilde\eta,\epsilon_c) = 
\frac{U \tilde\eta}
{\Delta E_{{\kappa^{\prime}_{2}\kappa_{2}\kappa^{\prime}_{1}\kappa_{1}}} 
- \epsilon_c} \ .
\label{hetaint}
\end{eqnarray}
Substituting the above expression into 
$\langle \tilde{H} \tilde{O}_{i} \rangle_{\rm HB}$, 
$\langle \tilde{O}^{\dagger}_{i}\tilde{H}\tilde{O}_{i} 
\rangle_{\rm HB}$, 
and 
$\langle \tilde{O}^{\dagger}_{i}\tilde{O}_{i} \rangle_{\rm HB}$, 
we have the forms such as 
$\langle \tilde{H} \tilde{O}_{i} \rangle_{\rm HB} = \langle
\tilde{O}^{\ast}_{i} \tilde{H} \rangle_{\rm HB} = 
\tilde{A} U^{2} \tilde{\eta}$,  
$\langle \tilde{O}^{\dagger}_{i}\tilde{H}\tilde{O}_{i}\rangle_{\rm HB} 
= \tilde{B} U^{2} \tilde{\eta}^{2}$, and 
$\langle \tilde{O}^{\dagger}_{i}\tilde{O}_{i} \rangle_{\rm HB} 
= \tilde{C} U^{2} \tilde{\eta}^{2}$.
Minimizing the energy $\epsilon_{c}$ with respect to 
$\tilde{\eta}$, we obtain
\begin{eqnarray}
\tilde{\eta}= \frac{-\tilde{B} + 
\sqrt{\tilde{B}^2+4\tilde{A}^2\tilde{C}U^2}}{2\tilde{A}\tilde{C}U^2} \ .
\label{hbetil}
\end{eqnarray}

The total energy should be obtained by taking the configurational 
average as 
\begin{eqnarray}
\langle H \rangle=\overline{ \langle H \rangle}_{\rm HB} + 
N \overline{\epsilon_c}\ .
\label{htote}
\end{eqnarray}
The HB ground-state energy is given by 
\begin{eqnarray}
\overline{ \langle H \rangle}_{\rm{HB}} &=& n\mu + 
\sum_{\sigma} \int^{0}_{-\infty}  
\epsilon \,\, \overline{\rho_{i\sigma}(\epsilon)} \, d\epsilon 
\nonumber \\
& & - (\overline U- \widetilde U) \overline{\langle \hat{n}_{i\uparrow} 
\rangle_{\rm HB} \langle \hat{n}_{i\downarrow} \rangle}_{\rm HB}
-\widetilde U  (\overline{n_{i\uparrow}\langle \hat{n}_{i\downarrow} 
\rangle}_{\rm HB}
+ \overline{n_{i\downarrow}\langle \hat{n}_{i\uparrow} 
\rangle}_{\rm HB})\ .
\label{hehb}
\end{eqnarray}
Here $\rho_{i\sigma}(\epsilon)$ is the local density of states (LDOS). 
It is obtained from the one-electron Green function, 
\begin{eqnarray}
\rho_{i\sigma}(\epsilon)= -\dfrac{1}{\pi}\, \rm {Im} \, 
\it G_{ii\sigma}(z) \ ,
\end{eqnarray}
and the Green function $G_{ii\sigma}(z)$ is defined by
Eq. (\ref{aagii}), in which   
$(\boldsymbol{H}_{\sigma})_{ij}$ has been replaced by 
the one-electron Hamiltonian 
matrix  for the HB Hamiltonian (\ref{hhb});  
$(\boldsymbol{H}_{\sigma})_{ij}=(\epsilon_{0}-\mu
+\overline U\langle \hat{n}_{i-\sigma} \rangle_{\rm HB}
+\widetilde U n_{i-\sigma}) \delta_{ij}
+ t_{i j} (1-\delta_{ij})$.
The average electron number $\langle \hat{n}_{i\sigma}\rangle_{\rm HB}$ 
is given by the LDOS as
\begin{eqnarray}
\langle \hat{n}_{i\sigma}\rangle_{\rm HB}=\int f(\epsilon) 
\rho_{i\sigma}(\epsilon) \, d\epsilon  \ .
\end{eqnarray} 

Since the HB Hamiltonian contains a random potential and the energy 
$\overline{ \langle H \rangle}_{\rm{HB}}$
is given by the LDOS, we can calculate the ground-state energy by means
of the alloy-analogy approximation, ({\it i.e.}, the CPA) as 
explained in the last subsection.  
In the CPA, we replace the random potentials at
the surrounding sites with a coherent potential $\Sigma_{\sigma}(z)$.  
The on-site impurity Green function is obtained as follows.
\begin{eqnarray}
G_{ii\sigma}(z)=\frac{1}{F_{\sigma}(z)^{-1}-\epsilon_0
+\mu-\overline{U}\langle \hat{n}_{i-\sigma}\rangle_{\rm HB} - 
\widetilde U n_{i-\sigma}+\Sigma_\sigma(z)} \ .
\end{eqnarray}
Here $F_{\sigma}(z)$ is the coherent Green function given by
Eq. (\ref{acohg2}). 

The self-consistent condition to determine the coherent potential
$\Sigma_{\sigma}(z)$ is given by Eq. (\ref{acpa}).  However,
$G_{ii\sigma}(z)$ for the HB potential fully depends on the 4 local
configurations $\alpha = (n_{i\uparrow}, n_{i\downarrow})$ via the
Hartree-Fock type potential 
$\overline{U}\langle \hat{n}_{i-\sigma}\rangle_{\rm HB}$
in the denominator.  Thus the configurational average of 
$G_{00\sigma}(z)$ is given by
\begin{eqnarray}
\overline{G_{00\sigma}(z)}=\sum_{\alpha} P_\alpha 
G^{\alpha}_{00\sigma}(z) \ .
\label{hgf}
\end{eqnarray}
Here  $P_\alpha$ is the probability when taking a configuration 
$\alpha$.
Note that instead of the configurations 
$\alpha= \, 00, \, 10, \, 01, \, 11 $, one can make use of an   
alternative notation $\nu=0$ (empty on a site), 
$1\uparrow$ (occupied by an electron with spin $\uparrow$ ),
$1\downarrow$ (occupied by an electron with spin $\downarrow$ ), and 
$2$ (occupied by 2 electrons). 
In this notation, we can express $P_\alpha$ as $P_0$, $P_{1\uparrow}$, 
$P_{1\downarrow}$,  and $P_2$. 
The impurity Green functions $G^{\alpha}_{00\sigma}(z)$ are given as
follows. 
\begin{eqnarray}
{G^{00}_{00\sigma}(z)}=\frac{1}{F_{\sigma}(z)^{-1}-\epsilon_0+\mu - 
\overline{U}\langle \hat{n}_{-\sigma}\rangle_{00}+\Sigma_\sigma(z)}\ ,
\label{himgf0}
\end{eqnarray}
\begin{eqnarray}
{G^{10}_{00\uparrow}(z)}=\frac{1}{F_{\sigma}(z)^{-1}-\epsilon_0 + 
\mu-\overline{U}\langle \hat{n}_{\downarrow}\rangle_{10} + 
\Sigma_\sigma(z)} \ ,
\end{eqnarray}
\begin{eqnarray}
{G^{10}_{00\downarrow}(z)}=\frac{1}{F_{\sigma}(z)^{-1}-\epsilon_0 + 
\mu-\overline{U}\langle \hat{n}_{\uparrow}\rangle_{10}-\widetilde U + 
\Sigma_\sigma(z)} \ ,
\end{eqnarray}
\begin{eqnarray}
{G^{01}_{00\uparrow}(z)}=\frac{1}{F_{\sigma}(z)^{-1}-\epsilon_0 + 
\mu-\overline{U}\langle \hat{n}_{\downarrow}\rangle_{01}-\widetilde U + 
\Sigma_\sigma(z)} \ ,
\end{eqnarray}
\begin{eqnarray}
{G^{01}_{00\downarrow}(z)}=\frac{1}{F_{\sigma}(z)^{-1}-\epsilon_0 + 
\mu-\overline{U}\langle \hat{n}_{\uparrow}\rangle_{01} + 
\Sigma_\sigma(z)} \ ,
\end{eqnarray}
\begin{eqnarray}
{G^{11}_{00\sigma}(z)}=\frac{1}{F_{\sigma}(z)^{-1}-\epsilon_0 + 
\mu-\overline{U}\langle \hat{n}_{-\sigma}\rangle_{11}-\widetilde U + 
\Sigma_\sigma(z)} \ ,
\label{himgf11}
\end{eqnarray}
and  the electron number for a given configuration $\alpha$ in the
denominators is given by
\begin{eqnarray}
\langle \hat{n}_{\sigma}\rangle_{\alpha}= \int  f (\epsilon) 
\rho^{\alpha}_{\sigma}(\epsilon)\, d \epsilon  \ ,
\label{hbnalpha}
\end{eqnarray}
\begin{eqnarray}
\rho^{\alpha}_{\sigma}(\epsilon) = -\dfrac{1}{\pi}\, 
{\rm Im} \, G^{\alpha}_{00\sigma}(z) \ .
\label{hdoshb}
\end{eqnarray}

The above expressions mean that the electron numbers 
$\langle \hat{n}_{\sigma}\rangle_{\alpha}$ have to be solved 
self-consistently for a given configuration with probabilities 
$\{ P_{\alpha} \}$ and for an effective medium $\Sigma_{\sigma}(z)$.  
The latter is obtained from the CPA equation (\ref{acpa}). 

The third and last terms at the r.h.s. of Eq. (\ref{hehb}) are
calculated in the SSA as follows.
\begin{eqnarray}
\overline{\langle \hat{n}_{i \uparrow} \rangle_{\rm HB} 
\langle \hat{n}_{i\downarrow} \rangle}_{\rm HB} = \sum_{\alpha} 
P_\alpha \langle \hat{n}_{ \uparrow} \rangle_{\alpha} 
\langle \hat{n}_{\downarrow}\rangle_{\alpha} \ ,
\label{hdblhb}
\end{eqnarray} 
\begin{eqnarray}
\sum_{\sigma} \overline{n_{i\sigma} 
\langle \hat{n}_{i -\sigma} \rangle}_{\rm HB} = 
\sum_{\sigma} \sum_{\alpha} P_\alpha
\, n^{\alpha}_{\sigma} \langle \hat{n}_{ -\sigma} \rangle_{\alpha} \ . 
\label{hp2f} 
\end{eqnarray} 
Here  $n^{\alpha}_{\uparrow} = 0, 1, 0, 1$ and 
$n^{\alpha}_{\downarrow} = 0, 0, 1, 1$ for 
$\alpha = 00, 10, 01, 11$, respectively.

The on-site probability satisfies the sum rule 
$P_0+ P_{1\uparrow}+P_{1\downarrow}+P_2=1$,  and 
the probability of finding an electron with spin 
$\uparrow(\downarrow)$ on a site is given by 
$P_{\uparrow(\downarrow)}=P_{1\uparrow(1\downarrow)}+P_2$. 
Therefore, $P_0, \, P_{1\uparrow}, \, \rm {and} \,P_{1\downarrow}$ 
are given by the probability $P_2$ in the paramagnetic state.

An approximate form of $P_{2}$ for the hybrid wavefunction is derived
as follows~\cite{pat13-2}.  
We have two kinds of approximate expressions for the
operator $\hat n_\uparrow \hat n_\downarrow$ according to the
alloy-analogy (AA) and Hartree-Fock (HF) approximation.
\begin{eqnarray}
\hat n_\uparrow \hat n_\downarrow \approx n_\uparrow  \hat n_\downarrow
+n_\downarrow  \hat n_\uparrow 
-n_\uparrow n_\downarrow\, \, \, \, \, \, (\rm AA) \ ,
\label{happha}
\end{eqnarray}
\begin{eqnarray}
\hat n_\uparrow \hat n_\downarrow \approx \hat n_\uparrow \langle \hat
 n_\downarrow\rangle_{\rm HB}
+\hat n_\downarrow \langle \hat n_\uparrow \rangle_{\rm HB}
-\langle \hat n_\uparrow \rangle_{\rm HB} 
\langle \hat n_\downarrow \rangle_{\rm HB}  \, \, \, \, \, \, (\rm HF)\ .
\label{happhh}
\end{eqnarray}
In the HB scheme, we superpose the above expressions with 
the weights $w$ and $1-w$, respectively.  
Taking the quantum mechanical and configurational average, we obtain an
approximate form of 
$P_{2} \,(=\overline{\langle \hat{n}_\uparrow \hat{n}_\downarrow 
\rangle})$.
Then, we have the term 
$w \,\overline{ n_\uparrow n_\downarrow} + 
(1-w) \overline{\langle \hat n_\uparrow \rangle_{\rm HB} \langle \hat
n_\downarrow \rangle}_{\rm HB}$ at the r.h.s., which may be again 
regarded as the probability $P_{2}$ in the HB scheme.  
Thus we obtain an approximate form of $P_{2}$ as follows.
\begin{eqnarray}
P_2 =  \frac{1}{2} w \, ( 
\overline {n_\uparrow \langle \hat{n}_\downarrow\rangle}_{\rm HB}
+ \overline {n_\downarrow \langle  \hat{n}_\uparrow \rangle}_{\rm HB})
+(1-w) \, \overline{\langle  \hat{n}_\uparrow \rangle_{\rm HB} 
\langle \hat{n}_\downarrow \rangle}_{\rm HB} \ .
\label{app2}
\end{eqnarray}
Since the r.h.s. of Eq. (\ref{app2}) is given by Eqs. (\ref{hdblhb}) 
and (\ref{hp2f}), we can self-consistently obtain the probabilities 
$\{ P_{\alpha} \}$.

Finally, the correlation energy $\overline{\epsilon}_{c}$ in
Eq. (\ref{htote}) is given as
\begin{eqnarray}
\overline{\epsilon_c}=\sum_{\alpha} P_\alpha \,  {\epsilon_{c\alpha}} \,  .
\label{hecbar}
\end{eqnarray} 
Here $\epsilon_{c\alpha}$ denotes the correlation energy for a given 
on-site configuration $\alpha$.
\begin{eqnarray}
 {\epsilon_{c\alpha}} = \Big[\dfrac{-\langle
 \tilde{O}^{\dagger}_{i}\tilde{H}\rangle_{\rm HB} -
\langle \tilde{H} \tilde{O}_{i} \rangle_{\rm HB} + 
\langle \tilde{O}^{\dagger}_{i}\tilde{H}\tilde{O}_{i}\rangle_{\rm HB}}
{1 + \langle \tilde{O}^{\dagger}_{i}\tilde{O}_{i} \rangle_{\rm HB}} 
\Big ]_\alpha \ .
\label{heca}
\end{eqnarray}
The quantities $\langle \tilde{H} \tilde{O}_{i} \rangle_{\rm HB}$, 
$\langle \tilde{O}^{\dagger}_{i}\tilde{H}\tilde{O}_{i}\rangle_{\rm HB}$, 
and $\langle \tilde{O}^{\dagger}_{i}\tilde{O}_{i} \rangle_{\rm HB}$ 
are expressed by the LDOS for the HB Hamiltonian, therefore the
correlation energy $\epsilon_{c\alpha}$ is obtained from the LDOS
$\rho^{\alpha}_{\sigma}(\epsilon)$ in the single-site CPA.

The double occupation number is obtained from 
$\partial \langle H \rangle/\partial U$.  Making use of the SSA,
we obtain 
\begin{eqnarray}
\overline{\langle \hat{n}_{i\uparrow}\hat{n}_{i\downarrow} \rangle} = 
\overline{\langle \hat{n}_{i\uparrow} \rangle_{\rm HB} 
\langle \hat{n}_{i\downarrow} \rangle}_{\rm HB} 
+ \overline{\langle \hat{n}_{i\uparrow}\hat{n}_{i\downarrow} 
\rangle}_{\rm c} \, ,
\label{hbdbl0}
\end{eqnarray}
Here $\overline{\langle \hat{n}_{i\uparrow} \rangle_{\rm HB} 
\langle \hat{n}_{i\downarrow} \rangle}_{\rm HB}$
has been obtained in Eq. (\ref{hdblhb}), and the correlation 
correction $\overline{\langle \hat{n}_{i\uparrow}\hat{n}_{i\downarrow} 
\rangle}_{\rm c}$
is given by
\begin{eqnarray}
\overline{\langle \hat{n}_{i\uparrow}\hat{n}_{i\downarrow} 
\rangle}_{\rm c} = \sum_{\alpha} 
P_\alpha \langle \hat{n}_{i\uparrow}\hat{n}_{i\downarrow} 
\rangle_{\rm{c}\alpha} \, .
\label{hbdblc}
\end{eqnarray}
Here $\langle \hat{n}_{i\uparrow}\hat{n}_{i\downarrow} 
\rangle_{\rm{c}\alpha}$ 
is the correlation correction for a given configuration $\alpha$, and is
given by Eq. (\ref{dblavc}) in which the operator $\tilde{O}_{i}$ has 
been replaced by Eq. (\ref{hboi}) and the average 
$\langle \sim \rangle_{0}$ has been replaced by 
$\langle \sim \rangle_{\rm HB}$.

The momentum distribution $\langle n_{k\sigma} \rangle$ is 
obtained from  $\partial \langle H \rangle / \partial 
(\epsilon_k -\sigma h)$
as follows.
\begin{eqnarray}
\langle n_{k\sigma} \rangle = 
\overline{\langle n_{k\sigma} \rangle}_{\rm HB} 
+ \overline{\langle n_{k\sigma} \rangle}_{c}\ .
\label{mlahbnk}
\end{eqnarray}
Here $\overline{\langle n_{k\sigma} \rangle}_{\rm HB}$ is the momentum
distribution in the hybrid state.
\begin{eqnarray}
\overline{\langle n_{k\sigma} \rangle}_{\rm HB}= \int 
f (\epsilon) {\rho_{k\sigma}(\epsilon)}\, d \epsilon   \ ,
\label{hnkhb}
\end{eqnarray}
\begin{eqnarray}
{\rho_{k\sigma}(\epsilon)} =  
-\dfrac{1}{\pi}\, \rm {Im} \,\it F_{k\sigma} \ .
\label{hmdos}
\end{eqnarray}
The Green function in the momentum representation is given in the CPA
as follows.
\begin{eqnarray}
{F}_{k\sigma} = 
\frac{1}{z - \Sigma_{\sigma}(z)-\epsilon_k} \ .
\end{eqnarray}
Here $\epsilon_k$ is the eigenvalue of $t_{ij}$ with momentum $k$.
 
The correlation correction $\overline{\langle n_{k\sigma} \rangle}_{c}$
is given as follows.
\begin{eqnarray}
\overline{\langle n_{k\sigma} \rangle}_{c} = 
\sum_{\alpha} P_\alpha \langle n_{k\sigma} \rangle_{c\alpha} \ .
\label{hbnk}
\end{eqnarray}
Here $\langle n_{k\sigma} \rangle_{\rm {c}\alpha}$ is the correlation
correction for the configuration $\alpha$, and is given by the second
term at the r.h.s. of Eq. (\ref{nkave})
in which $\tilde{O}_{i}$ has been replaced by Eq. (\ref{hboi}) and 
$\langle \sim \rangle_{0}$ has been replaced by 
$\langle \sim \rangle_{\rm HB}$.

\subsection{MLA-HB in infinite dimensions}

\begin{figure}
\begin{center}
\includegraphics[scale=0.4,angle=-90]{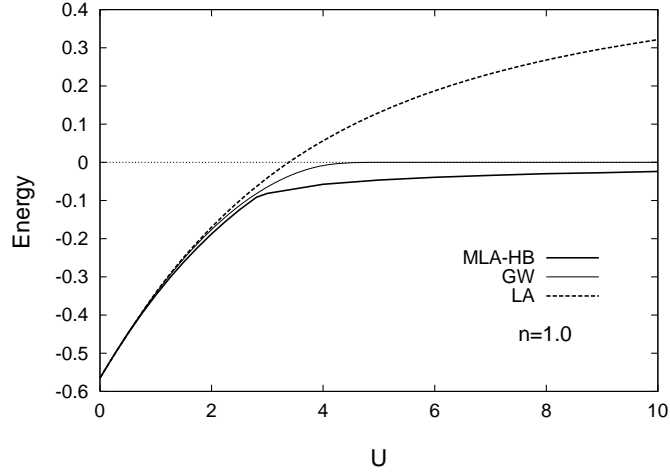}
\caption{\label{henergy} The energy vs Coulomb interaction energy
$U$ curves in the MLA-HB (solid curve), 
the GW (thin solid curve), and the LA (dotted curve) at half-filling 
($n=1.0$) (Ref. 47).}
\end{center}
\end{figure}

The MLA with HB wavefunction improves further the description of 
electron correlations in the
strongly correlated region.  
One can verify the fact by means of some numerical calculations in
infinite dimensions.
The ground state energy in the MLA-HB was obtained by
varying $w$ from 0 to 1 for each value of $U$.
Figure \ref{henergy} shows the ground-state
energy obtained by various methods on the hypercubic lattice 
in infinite dimensions at half filling.  
The energy in the LA monotonically
increases with increasing Coulomb interaction energy and becomes
positive beyond $U=3.4$ because it does not suppress sufficiently
the double occupancy in the strongly correlated region.
%
%
%
%

\begin{figure}
\begin{center}
\includegraphics[scale=0.4, angle=-90]{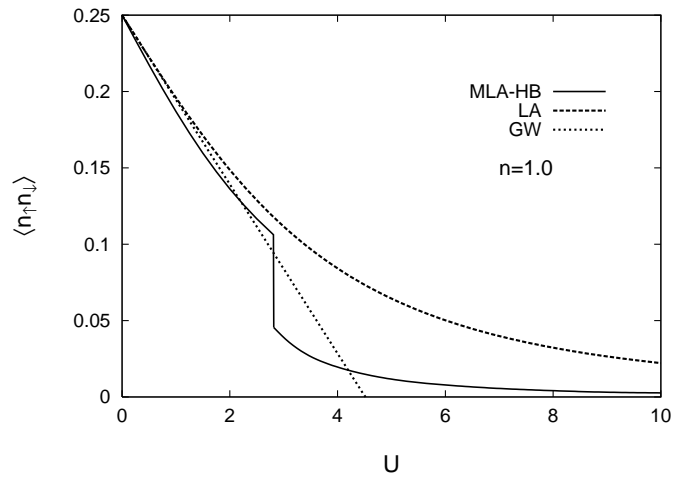}
\caption{\label{hdbl}The double occupation number  
$\langle n_{\uparrow}n_{\downarrow} \rangle$ 
vs Coulomb interaction energy $U$ curves in 
the MLA-HB (solid curve), 
the GW (dotted curve), and the LA (dot-dashed curve) 
at half-filling ($n=1.0$) (Ref. 47). }
\end{center}
\end{figure}

The ground-state energy in the GW is lower than that of the LA, and
approaches zero at $U_{c}({\rm GW})=4.51$ 
with increasing Coulomb interaction.
The Brinkman-Rice atomic state is realized beyond $U_{c}({\rm GW})$ 
(see Sec. 2.1).  
The ground-state
energy of the MLA-HB is the lowest among three wavefunctions over all
Coulomb interactions $U$.  
Note that there is a cusp in the energy vs $U$ curve at 
$U_{c}({\rm MLA})=2.81$. 
The Fermi-liquid ground state with $w=0$ is obtained below 
$U_{c}({\rm MLA})$, while the disordered local moment solution with
infinitesimal $w$ is stabilized beyond 
$U_{c}({\rm MLA})$~\cite{pat13-2}.
As shown in Fig. \ref{hdbl}, the double occupation number 
$\langle n_{\uparrow}n_{\downarrow} \rangle$ in the GW linearly
decreases with increasing $U$ according to Eq. (\ref{gdbl}).
In the case of the LA, it monotonically decreases according to
Eq. (\ref{ladbl}). 
The double occupation number in the MLA-HB is lower 
than that in the LA and GW in the weak Coulomb interaction regime
and jumps from 0.106 to 0.045 at $U_{c}({\rm MLA}) = 2.81$, 
indicating the metal-insulator transition.  
Beyond $U_{c}({\rm MLA})$, it again monotonically
decreases with increasing $U$.  Note that the double occupancy in the
MLA-HB remains finite in the strong $U$ regime as it should be, while
the GW gives the Brinkman-Rice atom, because the MLA takes into 
account the electron hopping from the atomic state.
%
%
%
%

\begin{figure}
\begin{center}
\includegraphics[scale=0.8]{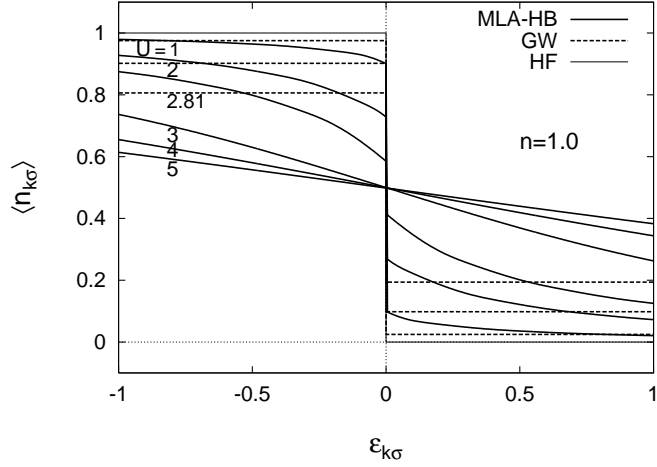}%
\caption{\label{hmomentum} The momentum distribution as a function of
energy  $\epsilon_{k\sigma}$ for various Coulomb interaction energy 
parameters $U=1.0,2.0,2.81,3.0,4.0$ and $5.0$ at half-filling. 
The MLA-HB: solid curves, the GW: dashed curves, and the HF: 
thin solid curve (Ref. 47).}
\end{center}
\end{figure}

The momentum distribution in the MLA-HB has the same behavior as the
MLA-HF in the metallic region; it decreases monotonically with 
increasing energy 
$\epsilon_{k\sigma}$ and shows a jump at the Fermi level, while it
disappears beyond $U_{c}({\rm MLA})$ as shown in Fig. \ref{hmomentum}.
With further increase of $U$, the curve becomes flatter.  These results
indicate that the MLA-HB improves upon the GW.  
Note that the distributions
in the GW are constant below and above the Fermi level 
irrespective of $U$ as discussed in Sec. 2.1.
The quasiparticle weight in the MLA-HB is the same as in the MLA-HF in
the metallic region (see Fig. \ref{figzu1}).  With the metal-insulator
transition at $U_{c}({\rm MLA})=2.81$, it disappears.  The existence of the
first-order transition at $U=U_{c}$ 
is in agreement with the result of the NRG~\cite{bulla01}, 
though $U_{c}$ in the NRG has not yet been published. 
%
%
%
%

\section{Towards the First-Principles MLA}

The momentum-dependent
local ansatz (MLA) wavefunction yields a reasonable description of
correlated electrons from the weak to strong Coulomb interaction regime,
and overcomes the limitations of 
the LA and the GW wavefunction, in particular, for the
description of physical quantities associated with the low-energy
excitations near the Fermi surface.  Therefore it is worthwhile 
to extend the MLA to the realistic system towards first-principles
calculations. 

Let us consider the first-principles LDA+U Hamiltonian which is
based on the tight-binding linear muffin-tin 
orbital~\cite{anisimov97,anisimov10}.

\begin{eqnarray}
H = H_{0} + H_{1} ,
\label{fphhat}
\end{eqnarray}
\begin{eqnarray}
H_{0} = \sum_{iL\sigma} \epsilon^{0}_{L} \, \hat{n}_{iL \sigma} 
+ \sum_{iL jL^{\prime} \sigma} t_{iL jL^{\prime}} \, 
a_{iL \sigma}^{\dagger} a_{jL^{\prime} \sigma} \ ,
\label{fph0}
\end{eqnarray}
\begin{eqnarray}
H_{1} & = & \sum_{i} 
\Big[ \sum_{m} U_{mm} \, \hat{n}_{ilm \uparrow} 
\hat{n}_{ilm \downarrow}  \nonumber \\
&  & \hspace{5mm} + {\sum_{m > m^{\prime}}} 
(U_{m m^{\prime}}-\frac{1}{2}J_{m m^{\prime}}) 
\hat{n}_{ilm} \hat{n}_{ilm^{\prime}} -
{\sum_{m > m^{\prime}}} J_{m m^{\prime}}   
\hat{\mbox{\boldmath$s$}}_{ilm} \cdot 
\hat{\mbox{\boldmath$s$}}_{ilm^{\prime}} 
\Big] \ . 
\label{fph1}
\end{eqnarray}
Here we assume a $d$ metal system with an atom per unit cell for
simplicity. $H_{0}$ and $H_{1}$ denote the non-interacting and
interacting parts, respectively. 
$\epsilon^{0}_{L}$ is an atomic level on site $i$ and 
orbital $L$, 
$t_{iL jL^{\prime}}$ is a transfer integral between orbitals $iL$ and 
$jL^{\prime}$. $L=(l,m)$ denotes $s$, $p$, and $d$ orbitals.
$a_{iL \sigma}^{\dagger}$ 
($a_{iL \sigma}$) is the creation (annihilation) operator for an
electron with orbital $L$ and spin $\sigma$ on site $i$, and 
$\hat{n}_{iL\sigma}=a_{iL \sigma}^{\dagger}a_{iL \sigma}$ is a charge
density operator for electrons with orbital $L$ and spin $\sigma$ on
site $i$. 
 
The inter-site Coulomb interactions are considered to be well screened
by the $s$ and $p$ band electrons, so that only the on-site 
Coulomb interactions between $d$ ($l=2$) electrons are taken into
account in Eq. (\ref{fph1}).
$U_{mm}$ ($U_{mm^{\prime}}$), and $J_{mm^{\prime}}$ denote 
the intra-orbital (inter-orbital)
Coulomb and exchange interactions, respectively.  
$\hat{n}_{ilm}$ ($\hat{\mbox{\boldmath$s$}}_{ilm}$) with $l=2$ is 
the charge (spin)
density operator for $d$ electrons on site $i$ and orbital $m$.
The atomic level $\epsilon^{0}_{L}$ in 
$H_{0}$ is calculated from the LDA atomic level
$\epsilon_{L}$ by subtracting the double counting potential; 
$\epsilon^{0}_{L} = \epsilon_{L} - \partial
E^{U}_{\rm LDA}/\partial n_{iL\sigma}$. 
Here $n_{iL\sigma}$ is the charge density at the ground state, 
$E^{U}_{\rm LDA}$ is a LDA functional for the intra-atomic Coulomb 
interactions~\cite{anisimov97,anisimov97-2}.

The Gutzwiller wavefunction (GW) has been extended 
to the case of a realistic Hamiltonian~\cite{bune97,bune97-2,bune98}.  
The wavefunction is constructed 
so as to reproduce the exact atomic states.  
We first solve the eigen value problem for the
atomic Hamiltonian in $H (= \sum_{i} H_{{\rm A}i})$.
The atomic Hamiltonian $H_{\rm A}$ on each site satisfies the eigen
value equation as follows.
\begin{eqnarray}
H_{\rm A} |\Gamma \rangle = E_{\Gamma} \, |\Gamma \rangle \, .
\label{fpatgamma}
\end{eqnarray}
Here we omitted the site index $i$ for simplicity.  $E_{\Gamma}$ denotes
the eigen value for the atomic eigen state $|\Gamma \rangle$ which is
obtained from the $2^{2M}$ atomic configuration states 
$\{ |I \rangle \}$ as 
$|\Gamma \rangle = \sum_{I} |I \rangle \, T_{I\Gamma}$, $M$ being the
number of orbitals in a site.

The atomic Hamiltonian is then expressed as 
\begin{eqnarray}
H_{\rm A} = \sum_{\Gamma} E_{\Gamma} \, \hat{m}_\Gamma \, ,
\label{fpath}
\end{eqnarray}
where $\hat{m}_\Gamma$ are the projection operators such that 
$\hat{m}_\Gamma = |\Gamma \rangle \langle \Gamma |$.
The Gutzwiller wavefunction is constructed as
\begin{eqnarray}
|\Psi_{\rm G} \rangle = P_{\rm G} \, |\phi \rangle = 
\big[ \prod_{i} P_{i} \big] \, |\phi \rangle \, .
\label{fpgw}
\end{eqnarray}
Here $|\phi \rangle$ is the Hartree-Fock wavefunction for the
Hamiltonian (\ref{fphhat}).

The local correlator $P_{i}$ in the wavefunction (\ref{fpgw}) is 
defined by
\begin{eqnarray}
P_{i} = 1 + \sum_{\Gamma} (\lambda_{i\Gamma} - 1) 
\hat{m}_{i\Gamma}\, .
\label{fppi}
\end{eqnarray}
Here we recovered the site index $i$.  $\lambda_{i\Gamma}$ denotes the
variational parameters.
The Gutzwiller wavefunction for the multiband Hamiltonian has been
applied to various correlated-electron systems such as Ni~\cite{bune03}
and Fe pnictides~\cite{wang10,schick12,bune12}.

The Gutzwiller wavefunction is constructed to reproduce well the atomic
regime.  But it does not reproduce the correct wavefunction in 
the weak Coulomb interaction limit.  
The MLA wavefunction is constructed to
reproduce the exact wavefunction in the weak interaction limit.
We rewrite the Hamiltonian as the sum of the Hartree-Fock Hamiltonian
$H_{\rm HF}$ and the residual interactions $H_{I}$.  
The latter is given by
\begin{eqnarray}
H_{I} & = & \sum_{i} 
\Big[ \sum_{m} U_{mm} \, O^{(0)}_{imm}  \nonumber \\
& & \hspace{10mm} + {\sum_{m > m^{\prime}}} 
(U_{m m^{\prime}} - \frac{1}{2}J_{m m^{\prime}}) 
O^{(1)}_{imm^{\prime}} -
{\sum_{m > m^{\prime}}} J_{m m^{\prime}}
O^{(2)}_{imm^{\prime}}   
\Big] \ . 
\label{fphres}
\end{eqnarray}
Here $O^{(0)}_{imm}$, $O^{(1)}_{imm^{\prime}}$, and 
$O^{(2)}_{imm^{\prime}}$ are the two-particle intra-orbital operators,
the charge-charge inter-orbital operators, and the spin-spin 
inter-orbital operators, respectively, which are defined as follows.
\begin{eqnarray}
O^{(0)}_{imm} \, & = & \ \delta\hat{n}_{ilm \uparrow} 
\delta\hat{n}_{ilm \downarrow}   \, , \\
O^{(1)}_{imm^{\prime}} & = & 
\ \delta\hat{n}_{ilm} \delta\hat{n}_{ilm^{\prime}}  \, , \\ 
O^{(2)}_{imm^{\prime}} & = & 
\ \delta\hat{\mbox{\boldmath$s$}}_{ilm} \cdot 
\delta\hat{\mbox{\boldmath$s$}}_{ilm^{\prime}} \ . 
\label{fpoi}
\end{eqnarray}

Applying the Rayleigh-Schr\"{o}dinger perturbation theory, we find that
the first-order correction $|\phi_{1} \rangle$ to the Hartree-Fock
wavefunction $|\phi \rangle$ is given by
\begin{eqnarray}
|\phi_{1} \rangle & = & - \sum_{i} \Big( \sum_{m} \tilde{O}^{(0)}_{imm} 
+ \sum_{m > m^{\prime}} \tilde{O}^{(1)}_{imm^{\prime}} +
{\sum_{m > m^{\prime}}} \tilde{O}^{(2)}_{imm^{\prime}}  \Big) 
\, |\phi \rangle \ . 
\label{fpphi1}
\end{eqnarray}
The two-particle operators 
$\tilde{O}^{(n)}_{iLL^{\prime}} \ (n=0, 1, 2)$ are
defined by
\begin{eqnarray}
\tilde{O}^{(n)}_{iLL^{\prime}} & = & \sum_{ \{ kn\sigma \} } 
\langle k^{\prime}_{2} n^{\prime}_{2}|iL \rangle 
\langle iL|k_{2} n_{2} \rangle
\langle k^{\prime}_{1} n^{\prime}_{1}|iL^{\prime} \rangle 
\langle iL^{\prime}|k_{1} n_{1} \rangle \nonumber \\
& & \hspace{5mm} \times 
\lambda^{(n)}_{LL^{\prime} \{ 2^{\prime}2\,1^{\prime}1 \}} 
\delta(a^{\dagger}_{k^{\prime}_{2}n^{\prime}_{2}\sigma^{\prime}_{2}} 
a_{k_{2}n_{2}\sigma_{2}})
\delta(a^{\dagger}_{k^{\prime}_{1}n^{\prime}_{1}\sigma^{\prime}_{1}} 
a_{k_{1}n_{1}\sigma_{1}}) \ .
\label{fpotilde}
\end{eqnarray}
Here $a^{\dagger}_{kn\sigma}$ ($a_{kn\sigma}$) is the creation
(annihilation) operator for an electron with momentum $k$,  band
index $n$, and spin $\sigma$.  They are related to those in the site
representation by  
$a_{kn\sigma} = \sum_{iL} a_{iL\sigma} \langle kn | iL \rangle$.
The momentum dependent coefficients 
$\lambda^{(n)}_{LL^{\prime} \{ 2^{\prime}2\,1^{\prime}1 \}}$ 
are given by 
\begin{eqnarray}
\lambda^{(0)}_{LL^{\prime} \{ 2^{\prime}21^{\prime}1 \}} & = & 
\eta_{L k^{\prime}_{2} n^{\prime}_{2} k_{2} n_{2} 
k^{\prime}_{1} n^{\prime}_{1} k_{1} n_{1}}
\delta_{LL^{\prime}} \delta_{\sigma^{\prime}_{2}\downarrow}
\delta_{\sigma_{2}\downarrow} \delta_{\sigma^{\prime}_{1}\uparrow}
\delta_{\sigma_{1}\uparrow} \, ,  \nonumber \\
\lambda^{(1)}_{LL^{\prime} \{ 2^{\prime}21^{\prime}1 \}} & = & 
\zeta^{(\sigma_{2}\sigma_{1})}_{LL^{\prime} k^{\prime}_{2} 
n^{\prime}_{2} k_{2} n_{2} k^{\prime}_{1} n^{\prime}_{1} k_{1} n_{1}}
\delta_{\sigma^{\prime}_{2}\sigma_{2}}
\delta_{\sigma^{\prime}_{1}\sigma_{1}} \, , \nonumber \\
\lambda^{(2)}_{LL^{\prime} \{ 2^{\prime}21^{\prime}1 \}} & = & 
\sum_{\sigma} \xi^{(\sigma)}_{LL^{\prime} k^{\prime}_{2} n^{\prime}_{2} 
k_{2} n_{2} k^{\prime}_{1} n^{\prime}_{1} k_{1} n_{1}}
\delta_{\sigma^{\prime}_{2} -\sigma} \delta_{\sigma_{2} \sigma}
\delta_{\sigma^{\prime}_{1}\sigma} 
\delta_{\sigma_{1} -\sigma}    \nonumber \\
& &  \hspace{5mm} + \frac{1}{2} \sigma_{1} \sigma_{2} 
\, \xi^{(\sigma_{2}\sigma_{1})}_{LL^{\prime} k^{\prime}_{2} 
n^{\prime}_{2} k_{2} n_{2} k^{\prime}_{1} n^{\prime}_{1} k_{1} n_{1}} 
\delta_{\sigma^{\prime}_{2}\sigma_{2}}
\delta_{\sigma^{\prime}_{1}\sigma_{1}} \, .  
\label{fplambda}
\end{eqnarray}

Finally, we obtain the MLA wavefunction for the realistic Hamiltonian 
(\ref{fphhat}) as follows.
\begin{eqnarray}
|\Psi_{\rm MLA} \rangle & = & \Big[ \prod_{i} 
\Big( 1 - \sum_{m} \tilde{O}^{(0)}_{imm} 
- \sum_{m > m^{\prime}} \tilde{O}^{(1)}_{imm^{\prime}} -
{\sum_{m > m^{\prime}}} \tilde{O}^{(2)}_{imm^{\prime}}  \Big) \Big] 
\, |\phi \rangle \ . 
\label{fpmlawf}
\end{eqnarray}
Here $\tilde{O}^{(0)}_{imm}$,\ $\tilde{O}^{(1)}_{imm^{\prime}}$,\, 
and $\tilde{O}^{(2)}_{imm^{\prime}}$ are the intra-orbital correlators,
the inter-orbital charge-charge correlators, and the inter-orbital
spin-spin correlators.  \\
$\eta_{L k^{\prime}_{2} n^{\prime}_{2} k_{2} n_{2} k^{\prime}_{1}
n^{\prime}_{1} k_{1} n_{1}}$, \ \ 
$\zeta^{(\sigma_{2}\sigma_{1})}_{LL^{\prime} k^{\prime}_{2} 
n^{\prime}_{2} k_{2} n_{2} k^{\prime}_{1} n^{\prime}_{1} k_{1} 
n_{1}}$, \ \ 
$\xi^{(\sigma)}_{LL^{\prime} k^{\prime}_{2} n^{\prime}_{2} 
k_{2} n_{2} k^{\prime}_{1} n^{\prime}_{1} k_{1} n_{1}}$, \ \  and \\  
$\xi^{(\sigma_{2}\sigma_{1})}_{LL^{\prime} k^{\prime}_{2} 
n^{\prime}_{2} k_{2} n_{2} k^{\prime}_{1} n^{\prime}_{1} 
k_{1} n_{1}}$ 
in these operators are the variational parameters.
The correlation energy $\epsilon_{c}$ 
is given by Eq. (\ref{mlaec}) with the
operator $\tilde{O}_{i}$ replaced by 
$\sum_{m} \tilde{O}^{(0)}_{imm} 
+ \sum_{m > m^{\prime}} \tilde{O}^{(1)}_{imm^{\prime}} +
{\sum_{m > m^{\prime}}} \tilde{O}^{(2)}_{imm^{\prime}}$.

Solving the self-consistent equations obtained from the stationary
condition $\delta \epsilon_{c} = 0$, we find approximate forms of the
variational parameters~\cite{kake14} 
which correspond to Eq. (\ref{etaint}). 
\begin{eqnarray}
\eta_{L k^{\prime}_{2} n^{\prime}_{2} k_{2} n_{2} k^{\prime}_{1}
n^{\prime}_{1} k_{1} n_{1}} = 
\dfrac{U_{mm} \, \tilde{\eta}_{m}}
{\Delta E_{k^{\prime}_{2} n^{\prime}_{2} \downarrow 
k_{2} n_{2} \downarrow k^{\prime}_{1}
n^{\prime}_{1} \uparrow k_{1} n_{1} \uparrow} - \epsilon_{\rm c}} \ ,
\label{fpetaip}
\end{eqnarray}
\begin{eqnarray}
\zeta^{(\sigma\sigma^{\prime})}_{LL^{\prime} k^{\prime}_{2} 
n^{\prime}_{2} k_{2} n_{2} k^{\prime}_{1} n^{\prime}_{1} k_{1} 
n_{1}} = 
\dfrac{(U_{mm^{\prime}} - J_{mm^{\prime}}/2) 
\, \tilde{\zeta}^{(\sigma\sigma^{\prime})}_{mm^{\prime}}}
{\Delta E_{k^{\prime}_{2} n^{\prime}_{2} \sigma k_{2} n_{2} 
\sigma k^{\prime}_{1}
n^{\prime}_{1} \sigma^{\prime} k_{1} n_{1} \sigma^{\prime} } 
- \epsilon_{\rm c}} \ ,
\label{fpzetaip}
\end{eqnarray}
\begin{eqnarray}
\xi^{(\sigma)}_{LL^{\prime} k^{\prime}_{2} 
n^{\prime}_{2} k_{2} n_{2} k^{\prime}_{1} n^{\prime}_{1} k_{1} 
n_{1}} = 
\dfrac{J_{mm^{\prime}} 
\, \tilde{\xi}^{(\sigma)}_{mm^{\prime}}}
{\Delta E_{k^{\prime}_{2} n^{\prime}_{2} -\sigma k_{2} n_{2} 
\sigma k^{\prime}_{1}
n^{\prime}_{1} \sigma k_{1} n_{1} -\sigma } 
- \epsilon_{\rm c}} \ ,
\label{fpxit}
\end{eqnarray}
\begin{eqnarray}
\xi^{(\sigma\sigma^{\prime})}_{LL^{\prime} k^{\prime}_{2} 
n^{\prime}_{2} k_{2} n_{2} k^{\prime}_{1} n^{\prime}_{1} k_{1} 
n_{1}} = 
\dfrac{J_{mm^{\prime}} 
\, \tilde{\xi}^{(\sigma\sigma^{\prime})}_{mm^{\prime}}}
{\Delta E_{k^{\prime}_{2} n^{\prime}_{2} \sigma k_{2} n_{2} 
\sigma k^{\prime}_{1}
n^{\prime}_{1} \sigma^{\prime} k_{1} n_{1} \sigma^{\prime} } 
- \epsilon_{\rm c}} \ .
\label{fpxil}
\end{eqnarray}
Here $\tilde{\eta}_{m}$, 
$\tilde{\zeta}^{(\sigma\sigma^{\prime})}_{mm^{\prime}}$, 
$\tilde{\xi}^{(\sigma)}_{mm^{\prime}}$, and 
$\tilde{\xi}^{(\sigma\sigma^{\prime})}_{mm^{\prime}}$ are new 
variational parameters.  
$\Delta E_{k^{\prime}_{2} n^{\prime}_{2} \sigma k_{2} n_{2} 
\sigma k^{\prime}_{1}
n^{\prime}_{1} \sigma^{\prime} k_{1} n_{1} \sigma^{\prime} } = 
\epsilon_{k^{\prime}_{2} n^{\prime}_{2} \sigma^{\prime}_{2}} - 
\epsilon_{k_{2} n_{2} \sigma_{2}} - 
\epsilon_{k^{\prime}_{1} n^{\prime}_{1} \sigma^{\prime}_{1}} - 
\epsilon_{k_{1} n_{1} \sigma_{1}}$
is the two-particle excitation energy.

The realistic multi-band Hamiltonian and its wavefunction contains new
physics: (1) orbital-dependent suppression of charge fluctuations, (2)
formation of atomic magnetic moments due to Hund's rule couplings,
(3) intra-atomic instabilities such as the high-spin to the low-spin
transition, (4) band-dependent quasiparticle weight ({\it i.e.},
effective masses), (5) orbital selective metal-insulator transition.
Implementation of the first-principles MLA calculations is left 
for future investigations~\cite{kake14}.

\section{Summary}

We have presented recent progress in the development 
of the local ansatz wavefunction with
momentum dependent variational parameters, {\it i.e.}, the MLA.
The MLA wavefunction takes into account the Hilbert space expanded by
the two-particle operators of the residual Coulomb interactions, as in
the local ansatz (LA) function, but the amplitudes for the 
two-particle excited states in the momentum
representation are assumed to be momentum dependent so as to be 
exact in the weak Coulomb interaction limit.  Consequently, the 
two-particle operators $\{ \eta_{i} O_{i} \}$ in the LA are replaced
by a new set of operators $\{ \tilde{O}_{i} \}$ with momentum-dependent
variational parameters 
$\{ \eta_{k^{\prime}_{2}k_{2}k^{\prime}_{1}k_{1}} \}$ in the MLA.
By making use of the variational principle, we determine the best
amplitudes $\eta_{k^{\prime}_{2}k_{2}k^{\prime}_{1}k_{1}}$
for two-particle excited states and 
again project those states with the best amplitudes onto the local 
subspace. 

We have demonstrated that the MLA improves upon the LA in terms of 
energy irrespective
of the electron number and the Coulomb interaction energy, and 
more strongly 
suppresses the double occupation number as compared with the LA.  
In particular, the momentum
distribution functions show a momentum dependence, while 
those in the LA and the Gutzwiller wavefunction (GW) show a flat
behavior below and above the Fermi level.
Furthermore, the quasiparticle weight vs Coulomb interaction curve is
close to the NRG result, while those in the LA and the GW strongly
deviates from the NRG curve.

One can improve the MLA by changing the starting wavefunction from the
Hartree-Fock (HF) wavefunction to the hybrid (HB) one.  
The HB wavefunction is the ground state for the HB
Hamiltonian which is a superposition of the
Hartree-Fock Hamiltonian and the alloy-analogy (AA) one.  
The HF wavefunction is a good starting wavefunction 
in the weakly correlated regime, while the AA wavefunction is 
more suitable in the strongly correlated regime. 
One can choose the best HB wavefunction by controlling the 
weighting parameter in the HB Hamiltonian.  
The MLA with the HB wavefunction is
applicable to both the weak and the strong Coulomb interaction 
systems. 

The MLA-HB yields a lower ground-state energy than the LA and
GW, and causes the first-order metal-insulator transition, at which 
point the double occupation number jumps as a function of the Coulomb 
interaction strength.  
The double occupation number in the insulating regime remains
finite as it should be in infinite dimensions, 
while it vanishes in the GW because of
the appearance of the Brinkman-Rice atomic state.  The momentum
distribution function shows a clear momentum dependence in both the
metallic and the insulating regions.  These facts indicate that the
MLA-HB well describes correlated electrons in high dimensions.  
We have also extended the MLA
to the realistic system.  The first principles MLA can describe the
charge-charge correlations and the spin-spin correlations
between electrons on the different orbitals, 
in addition to the intra-orbital correlations.
It is useful for understanding the correlated electrons in 
real systems.  

The MLA wavefunction method presented here is limited to the single-site
approximation.  Inclusion of nonlocal correlations is desired 
to describe 
magnetism, the metal-insulator transition, and the frustrated electrons 
in low dimensional systems.  There one needs to introduce 
explicitly the nonlocal operators such as
$\tilde{O}_{ij} = \sum_{k_{1}k_{2}k^{\prime}_{1}k^{\prime}_{2}} 
\langle k^{\prime}_{1}|i \rangle \langle i|k_{1} \rangle 
\langle k^{\prime}_{2}|j \rangle \langle j|k_{2} \rangle
\eta_{k^{\prime}_{2}k_{2}k^{\prime}_{1}k_{1}} 
\delta(a^{\dagger}_{k^{\prime}_{2}\downarrow}a_{k_{2}\downarrow})
\delta(a^{\dagger}_{k^{\prime}_{1}\uparrow}a_{k_{1}\uparrow})$ 
with momentum-dependent variational parameters 
$\eta_{k^{\prime}_{2}k_{2}k^{\prime}_{1}k_{1}}$.  
Extension of the MLA to the nonlocal case is also left as 
future work.

\section*{Acknowledgments}

Professor Martin C. Gutzwiller passed away on 3 March 2014.
The authors would like to express their sincere thanks to him 
for his continuous interest in the MLA and for his encouragement to them.

\end{document}